\documentclass[aps,pra,twocolumn,superscriptaddress,amsmath,floatfix,nofootinbib,10pt]{revtex4-2}
\usepackage{amsmath,amssymb,amsthm,bm,amsfonts,mathrsfs,bbm}
\usepackage{txfonts}
\usepackage[T1]{fontenc}
\usepackage{natbib}
\usepackage{dsfont}
\usepackage{graphicx}
\usepackage{bm}
\usepackage{ragged2e}
\usepackage[colorlinks=true,linkcolor=blue,citecolor=blue,urlcolor=blue]{hyperref}
\newcommand{\Tr}{\operatorname{Tr}} 

\usepackage{xspace}  
\usepackage{pgf,tikz}
\usepackage{xcolor}
\usepackage{physics}
\newtheorem{theorem}{Theorem}
\newtheorem{corollary}{Corollary}
\newtheorem{definition}{Definition}
\newtheorem{proposition}{Proposition}

\newtheorem{remark}{Remark}
\newtheorem{lemma}{Lemma}

\newcommand{\Z}{\mathbb{Z}}

\newcommand{\C}{\mathbb{C}}

\begin{document}
\title{Magic Moments: An Operational Approach of Certifying Nonstabilizerness from Few Copies}

\author{Kunal Sharma\textsuperscript{\#}}
\email{kunal.qic@gmail.com}
\affiliation{School of Physical Sciences, Indian Institute of Technology Goa, Ponda 403401, Goa, India}

\author{Sudip Chakrabarty\textsuperscript{\#}}
\email{sudip27042000@gmail.com}
\affiliation{S. N. Bose National Centre for Basic Sciences, Block JD, Sector III, Salt Lake, Kolkata 700 106, India}

\author{Saheli Mukherjee}
\email{mukherjeesaheli95@gmail.com}
\affiliation{S. N. Bose National Centre for Basic Sciences, Block JD, Sector III, Salt Lake, Kolkata 700 106, India}

\author{Bivas Mallick}
\email{bivasqic@gmail.com}
\affiliation{Statistics and Mathematics Unit, Indian Statistical Institute, 560059 Bangalore, India}

\author{Ananda G. Maity}
\email{anandamaity289@gmail.com}
\affiliation{School of Physical Sciences, Indian Institute of Technology Goa, Ponda 403401, Goa, India}

{\renewcommand{\thefootnote}{\#}\footnotetext{These authors contributed equally to this work.}}

\begin{abstract}
Magic is the key resource enabling universal fault-tolerant quantum computation beyond classically simulable stabilizer circuits. Certifying it, however, usually requires either full state tomography or an optimization over the set of stabilizer states, both of which rapidly become prohibitive as the system grows. Here, we introduce a moment-based criterion that certifies magic from a few moments of the discrete Wigner function, without fully reconstructing the distribution itself. For pure states of any number of qudits the criterion is necessary and sufficient, while for mixed states we construct a hierarchy of higher moments conditions that reaches the exact negativity threshold. We also derive lower bounds on sum negativity and mana from low-order moments, and use them to bound the negativity factor in quasiprobability simulation. Each moment can be evaluated as the expectation value of a two-outcome observable on a few copies of the state, measurable with a constant-depth Clifford circuit and a number of rounds independent of qudit number, providing efficient, experimentally accessible tools for certifying and quantifying nonstabilizer resources.
\end{abstract}
\maketitle

\section{Introduction}\label{s1}
Universal fault-tolerant quantum computation offers a promising route to information-processing capabilities beyond the reach of classical computers~\cite{Deutsch_2020}. Identifying the quantum resources underlying this potential advantage remains a fundamental challenge. Although entanglement is essential for many quantum information-processing tasks, it alone does not guarantee a computational advantage. Indeed, the Gottesman-Knill theorem establishes that circuits comprising stabilizer-state preparations, Clifford operations, and computational-basis measurements can be efficiently simulated classically, even when they generate highly entangled states~\cite{gottesman_1998}. Achieving universal and efficient quantum computation, in contrast, requires extending this classically simulable framework with suitable non-Clifford operations~\cite{Bravyi_2005,Hosseinynejad_2026}. The quantum resource associated with this departures, known as \textit{nonstabilizerness} or \textit{quantum magic}, therefore plays a central role in enabling quantum computational advantage~\cite{Bravyi_2005,Hosseinynejad_2026,veitch_2014,howard_2017}.

Beyond computation, nonstabilizerness has emerged as an important resource in quantum many-body physics and communication complexity, offering insights into quantum dynamics, resource generation, and the limits of classical information processing~\cite{Liu_2022,Falco_25,Chowdhury_2026}. For odd-prime-dimensional qudits, the discrete Wigner function provides a natural phase-space description, connecting negativity to contextuality and limitations of efficient classical simulation within the quasiprobability framework~\cite{gross_2006,veitch_2012,mari_2012,howard_2014}. These connections have motivated the development of several quantitative measures of nonstabilizerness, including the mana and relative entropy of magic~\cite{veitch_2014}, robustness of magic~\cite{howard_2017}, thauma~\cite{wang_2020}, and stabilizer R\'{e}nyi entropy (SRE)~\cite{leone_2022,wang_2023}. Although these measures capture complementary aspects of quantum magic, their experimental accessibility and computational evaluation remain significant challenges.

The difficulty stems from two fundamental bottlenecks. Experimentally, many existing approaches require extensive knowledge of the quantum state, with full tomography incurring a measurement overhead that scales exponentially with the number of qudits~\cite{haah_2017,odonnell_2016}. Computationally, the stabilizer polytope, defined as the convex hull of pure stabilizer states, has a rapidly growing number of vertices, making its exact characterization increasingly demanding. Consequently, determining whether a state belongs to this polytope or quantifying its distance from the stabilizer set can become computationally intractable even when complete tomographic information is available. These limitations often restrict practical methods to systems of limited size or dimension, or to states possessing specific symmetries or structures.

To circumvent these difficulties, several measurement-based approaches have been proposed to probe quantum magic without complete state reconstruction. Randomized-measurement protocols based on the second stabilizer R\'{e}nyi entropy enable magic quantification but suffer from sensitivity to experimental imperfections and require exponentially many measurements with increasing system size~\cite{oliviero_2022}. More robust entropy-based criteria have since been developed, although their efficient certification guarantees hold only under restricted entropy conditions and not for arbitrary mixed states~\cite{haug_2026}. Complementary methods exploit two-copy Bell-sampling statistics~\cite{haug_2023}, while multifractal-flatness approaches can require exponentially many samples to estimate stabilizer entropy to fixed accuracy~\cite{turkeshi_2023}. More recently, the reduced-stabilizer-polytope framework has enabled magic detection from a restricted set of Pauli measurements, together with lower bounds on the robustness of magic~\cite{varela2026predicting}. Despite these advances, experimentally accessible and scalable methods for detecting and quantifying magic in higher-dimensional quantum systems remain elusive. This raises a fundamental question: \textit{Can quantum magic be detected and quantitatively certified from limited measurement data, without full state tomography or computationally demanding optimization over the stabilizer polytope?}

In this work, we address this question affirmatively by introducing a moment-based framework for the operational detection and quantification of quantum magic in qudit systems. The use of moments has proven effective in detecting a variety of quantum resources \cite{PhysRevLett.121.150503,elben_2020,neven2021symmetry,PhysRevLett.127.060504,PhysRevA.109.022247,PhysRevA.111.032406,nzrc-8yrt,ffch-xyv2,lkcb-nqqb,tarabunga2026quantifying,chakrabarty2026operational, 828t-3czf,bvkj-2fbz,chakrabarty2026quantifyingmargenauhillnonclassicality,chakrabarty2026many}, motivating their application to the characterization of quantum magic. Working within the discrete Wigner representation, we establish a universal inequality involving low-order moments whose violation certifies Wigner negativity and hence nonstabilizerness. For pure states of arbitrarily many qudits of odd prime dimension, we prove that this criterion is both necessary and sufficient, invoking the discrete Hudson theorem~\cite{gross_2006}. For mixed states, we construct a hierarchy of higher-moment conditions that progressively approaches the exact threshold for Wigner negativity. Beyond detection, we derive two complementary lower bounds on the sum negativity and mana from different combinations of low-order moments. These bounds also provide operational estimates of the sampling overhead associated with quasiprobability-based classical simulations of Clifford circuits supplied with magic-state inputs~\cite{pashayan_2015}.

Importantly, our framework admits an experimentally accessible implementation. We show that every moment whose order is not a multiple of the local dimension can be estimated through measurements of dichotomic observables acting on a few copies of the state, implemented using constant-depth Clifford circuits. Each such moment can be estimated to additive accuracy $\varepsilon$, with confidence at least $1-\delta$, using $\mathcal{O}\bigl(\varepsilon^{-2}\log(1/\delta)\bigr)$ measurement rounds, independently of the number of qudits. Unlike full state tomography, whose cost grows exponentially with system size, our approach therefore has a sample complexity that is size-independent at fixed additive accuracy, and it requires no optimization over the stabilizer polytope. We further investigate moment estimation using single-copy randomized measurements and illustrate the detection and quantification criteria with representative qutrit and ququint states, including the strange state and its noisy variants. Together, these results establish magic moments as an operational tool for certifying nonstabilizerness and deriving quantitative resource bounds directly from experimentally accessible information.

\section{Results}\label{sec:results}
Throughout the paper, $n$ denotes the number of qudits, $d$ their odd prime local dimension, and $D=d^n$ the total dimension. On a single qudit, the shift and phase operators
satisfy $X\ket{j}=\ket{j+1}$ and $Z\ket{j}=\omega^j\ket{j}$, where
$\omega=e^{2\pi i/d}$ and arithmetic is modulo $d$. The displacement
operator at $(p,q)\in\mathbb Z_d^2$ is
\begin{equation}
T_{(p,q)}=\omega^{-2^{-1}pq}Z^pX^q,
\end{equation}
where $2^{-1}$ is the inverse of $2$ modulo $d$. For
$\mathbf u=(\mathbf p,\mathbf q)\in\mathbb Z_d^{2n}$, the corresponding
$n$-qudit displacement is $T_{\mathbf u}=\bigotimes_{\ell=1}^{n}T_{(p_\ell,q_\ell)}$.

The operators $\omega^a T_{\mathbf u}$, with $a\in\mathbb Z_d$, form
the Pauli group $\mathcal H_d$. The Clifford group is its normalizer,
\begin{equation}
\mathcal C_d=\{U\in\mathrm U(D):U\mathcal H_dU^\dagger=\mathcal H_d\}.
\end{equation}
A pure stabilizer state is the unique common $+1$ eigenstate of an
abelian Pauli subgroup of order $D$ that contains no nonidentity scalar.
The convex hull of pure stabilizer states is the stabilizer polytope.
States outside this polytope are magic states.

The phase-point operators are defined by~\cite{gross_2006}
\begin{equation}
A_{\mathbf 0}=\frac{1}{D}\sum_{\mathbf v}T_{\mathbf v},
\qquad
A_{\mathbf u}=T_{\mathbf u}A_{\mathbf 0}T_{\mathbf u}^{\dagger}.
\label{eq:A_u}
\end{equation}
Here $A_{\mathbf 0}\ket{\mathbf x}=\ket{-\mathbf x}$. The operators
satisfy $\Tr(A_{\mathbf u})=1$ and
$\Tr(A_{\mathbf u}A_{\mathbf v})=D\delta_{\mathbf u,\mathbf v}$.
The discrete Wigner function (DWF) is
\begin{gather}
W_\rho(\mathbf u)=\frac{1}{D}\Tr(A_{\mathbf u}\rho),
\qquad \sum_{\mathbf u}W_\rho(\mathbf u)=1,
\label{eq:DWF}
\end{gather}\\
and its moments are defined in Eq. \eqref{eq:magic_moment}. By the discrete Hudson theorem, Wigner negativity is equivalent to magic for pure states. For mixed states, Wigner negativity is sufficient to certify magic~\cite{gross_2006,veitch_2012}. The Supplementary Materials give further details of the formalism along with examples.

\subsection{Detecting magic from low-order Wigner moments}\label{s3}
We first introduce a moment-based framework for detecting nonstabilizerness, then show how to experimentally measure the required moments and use them to quantify Wigner negativity.

\begin{definition}\label{definition:magic_moment}
    For a quantum state $\rho\in \mathcal{D}(\C^d)$, with discrete Wigner function $W_{\rho}(\textbf{u})$ (with $\textbf{u}\in \Z_d\times\Z_d$ labels the points of the discrete phase-space), we define its $k$-th Wigner moment as
    \begin{equation}
     w_k :=\sum_{\textbf{u}=(0,0)}^{(d-1,\,d-1)}[W_{\rho}{(\textbf{u})}]^k,  \label{eq:magic_moment}
     \end{equation}
    where, $k\in\mathbb{N},\; \text{ and } k\geq 1$.
\end{definition}

\subsubsection{A necessary and sufficient criterion for detecting pure magic states}\label{subsec:faithful}
 
By the discrete Hudson theorem, a pure state is magic if and only if its Wigner function is negative at some phase-space point~\cite{gross_2006}. Checking this directly requires evaluating the Wigner function at every point, which is hard in general. We show that for pure states a single number built from the moments suffices, with no need for individual phase-space values. We first review three properties of the Wigner moments that hold for any number of qudits.
 
\begin{lemma}\label{lem:basic}
For every state $\rho$ of $n$ qudits of odd prime dimension $d$, the Wigner function and its moments satisfy
\begin{align}
\big|W_\rho(\mathbf u)\big| &\le \frac{1}{D} \qquad \text{for all } \mathbf u, \label{eq:Wbound}\\[2pt]
w_2 &= \frac{1}{D}\,\Tr\big[\rho^2\big], \label{eq:w2purity}\\[2pt]
|w_k| &\le \frac{w_2}{D^{\,k-2}} \qquad \text{for all } k\ge 2. \label{eq:wkbound}
\end{align}
\end{lemma}
 
The detailed proof is given in Methods~\ref{methods:proof_basic}. Eq.~\eqref{eq:w2purity} tells us that the second moment is simply the purity, up to the factor $D$. We therefore introduce rescaled moments whose magnitudes are bounded independently of the number of qudits,
\begin{align}
P = D\,w_2 = \Tr\big[\rho^2\big], \quad b = D^2\,w_3, \quad c = D^3\,w_4 . \label{eq:rescaled}
\end{align}
In these variables, Lemma~\ref{lem:basic} takes a compact form,
\begin{align}
|b|\le P, \qquad 0\le c\le P. \label{eq:rescaledbounds}
\end{align}
 
For a pure state the purity is one, so $P=1$ and this inequality reduces to a condition on $b$ alone. The bound $|b|\le P$ shows that $b$ can never exceed one. The central question is therefore whether the upper value $b=1$ is reserved for stabilizer states. The following theorem answers it.
 
\begin{theorem}\label{thm:faithful}
Let $\ket{\psi}$ be a pure state of $n$ qudits of odd prime dimension $d$. Then
\begin{align}
D^2\, w_3 \le 1, \label{eq:w3pure}
\end{align}
and equality holds if and only if $\ket{\psi}$ is a stabilizer state. Consequently,
\begin{align}
D^2w_3>1 \quad\Longleftrightarrow\quad \ket{\psi}\ \text{is a magic state}. \label{eq:faithful}
\end{align}
\end{theorem}
 
For a detailed proof see Methods~\ref{methods:proof_pure}. Theorem~\ref{thm:faithful} is one of the main detection result of this work. For pure states the moment criterion is not merely sufficient but exact, so no pure magic state can escape it, and it depends only on the third moment, regardless of the number of qudits.

The single number $D^2w_3$ has two further properties that make it a natural figure of merit. It does not change under Clifford unitaries, because these only permute the points of phase space~\cite{gross_2006} and every moment is a sum over all points. It is also multiplicative under tensor products, because the Wigner function of a product state is the product of the Wigner functions.

To make the certificates concrete, we use the strange state 
\begin{equation}
|\psi_1\rangle=\frac{\ket{1}-\ket{2}}{\sqrt{2}},
\label{eq: strange}
\end{equation}
and the ququint state
\begin{equation}
\ket{\psi_4}=\frac{\ket0-\ket1+\ket2+\ket3+\ket4}{\sqrt5}.
\label{eq:psi4}
\end{equation}
Their rescaled third moments are $b=0$ and $b=77/125=0.616$, respectively, compared with $b=1$ for every pure stabilizer state. The detailed discrete Wigner distributions and the original pure-state examples are provided in the Supplementary Information.


\subsubsection{Extension to mixed states}\label{subsec:mixed}
The proof of Theorem~\ref{thm:faithful} relies essentially on purity, and Hudson’s theorem does not extend to mixed states. The moment inequality below, however, requires only Wigner nonnegativity, so it applies to both pure and mixed states.

\begin{theorem} \label{thm:main}
If $\rho\in\mathcal{D}(\mathbb{C}^D)$ is a quantum state whose discrete Wigner function is nonnegative at every point of the discrete phase space, i.e., $W_{\rho}(\textbf{u})\geq0\;\forall\;\textbf{u}\in \Z_d^{2n}$ then 
\begin{equation}
   w_2^2\leq w_3 \label{eq:2_3_moment_condition}
\end{equation}
where, $w_2$ and $w_3$ are defined in Definition~\ref{definition:magic_moment}.
\end{theorem}

A detailed proof is given in Methods~\ref{methods:proof_mixed}. The condition holds for every non-negative, normalized DW, independent of the Hilbert-space dimension and the purity of the state, so any violation certifies Wigner negativity. For pure states, it recovers one direction of Theorem~\ref{thm:faithful}, which also proves the converse. To establish Theorem~\ref{thm:main} as a witness of magic for mixed states, we must show that every mixed stabilizer state has a non-negative Wigner function. We begin with the following lemma.

\begin{lemma}\label{lemma:mixed}
Let $\rho=\sum_k p_k|S_k\rangle\langle S_k|$ be a mixed stabilizer state in an odd prime or prime-power dimension, where $\{p_k\}$ forms a probability distribution and $|S_k\rangle$ are pure stabilizer states. Then, 
\begin{equation} W_\rho(\mathbf{u})\geq0, \qquad \forall\,\mathbf{u}\in\mathbb{Z}_d\times\mathbb{Z}_d. 
\end{equation}
\end{lemma}

The proof is given in Methods~\ref{methods:proof_stabilizer}. The above lemma implies that a violation of Theorem~\ref{thm:main} certifies a magic state. A nonnegative Wigner function, however, does not imply stabilizer-polytope membership for mixed states. The rescaled witness is $G\equiv P^2-b=D^2(w_2^2-w_3)$, which is nonpositive for every nonnegative Wigner function, so $G>0$ certifies magic.

\begin{figure*}[t]
        \centering
        \includegraphics[width=0.93\linewidth]{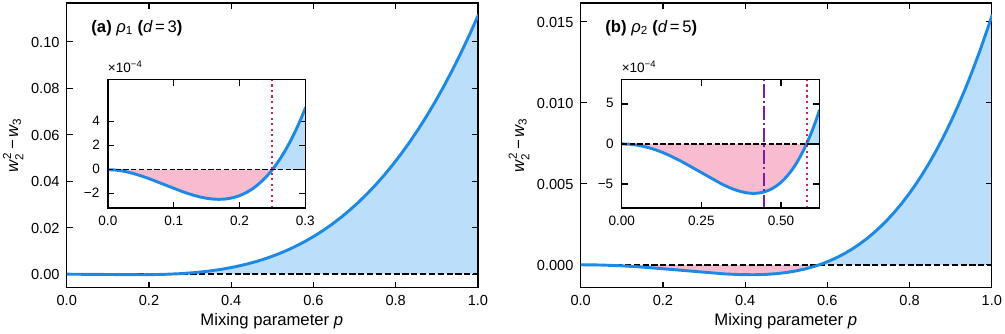}
        \caption{\justifying \small Analytical graphs of $(w_2^2-w_3)$ as a function of the mixing parameter $p$ for (a) the qutrit state $\rho_1$ ($d=3$) and (b) the ququint state $\rho_2$ ($d=5$). The blue curves represent the corresponding analytical expressions. Light blue and light pink shading indicate positive and negative values, respectively, with positive values certifying magic according to the criterion. The black dashed lines represent zero. The insets magnify the regions near $p=0$. In (a), the magenta dotted line marks $p=1/4$, where both the criterion and the Wigner function change sign. In (b), the purple dash-dotted line marks $p=1/\sqrt{5}$, above which the Wigner function becomes negative, while the magenta dotted line marks $p=(3+\sqrt{409})/40$, above which the criterion is violated.}
        \label{mixed_states_combined}
\end{figure*}

For illustration consider the two depolarized families
\begin{align}
\rho_1(p)&=p\ket \psi_1\bra \psi_1+(1-p)\mathbb I_3/3,\nonumber \\
\rho_2(p)&=p\ket{\psi_4}\bra{\psi_4}+(1-p)\mathbb I_5/5,
\label{eq:noisy_families}
\end{align}
with $0\leq p\leq1$. Their full Wigner distributions and moment calculations are provided in the Supplementary Information. For the noisy strange state,
\begin{equation}
w_2^2-w_3=\frac{p^2(p+2)(4p-1)}{81},
\label{eq:noisy_strange_witness}
\end{equation}
so the third-moment witness detects exactly the Wigner-negative region $p>1/4$. The fourth-moment witness has the same threshold. For the noisy ququint,
\begin{equation}
w_2^2-w_3=\frac{4p^2(20p^2-3p-5)}{3125},
\label{eq:noisy_ququint_witness}
\end{equation}
so the third-moment threshold is $(3+\sqrt{409})/40\simeq0.5806$, whereas Wigner negativity starts at $1/\sqrt5\simeq0.4472$. Figure~\ref{mixed_states_combined} displays this distinction.

\subsubsection{A complete hierarchy for mixed states}\label{subsec:hierarchy}

For mixed states the criterion of Theorem~\ref{thm:main} is sufficient but not necessary. The $d=5$ example of Sec.~\ref{subsec:mixed} makes this explicit, since the DWF of $\rho_2$ is negative for $p>1/\sqrt5\approx0.447$ while $w_2^2>w_3$ holds only for $p>0.581$. One may ask whether a better inequality in $w_2$ and $w_3$ alone exists. It does not. Any normalized distribution that is uniform on its support saturates $w_2^2=w_3$, so the Cauchy-Schwarz bound is already optimal at this order. The gap can only be closed with higher moments, and the following theorem shows how.

\begin{theorem}\label{thm:hankel}
For a state $\rho$ and an integer $m\ge1$ define the $(m+1)\times(m+1)$ matrix
\begin{align}
\big[H_m(\rho)\big]_{ij}=w_{i+j+1}, \qquad i,j=0,1,\dots,m, \label{eq:hankel}
\end{align}
with $w_1=1$. If $W_\rho(\mathbf u)\ge0$ for all $\mathbf u$, then $H_m(\rho)\succeq0$ for every $m$. Conversely, if $W_\rho$ takes $r$ distinct values and is negative somewhere, then $H_{r-1}(\rho)$ has a negative eigenvalue. Since $r\le D^2$, the conditions $H_m\succeq0$ with $m\le D^2-1$ are necessary and sufficient for a nonnegative DWF.
\end{theorem}

We provide the proof in Methods~\ref{methods:proof_hierarchy}. The first level is the matrix
\begin{align}
H_1=\begin{pmatrix}1&w_2\\ w_2&w_3\end{pmatrix}, \label{eq:H1}
\end{align}
whose determinant is $w_3-w_2^2$, so that $H_1\succeq0$ is exactly Theorem~\ref{thm:main}. Level $m$ requires the moments up to $w_{2m+1}$, each of which can be measured with the multicopy protocol of Sec.~\ref{sec:multicopy}. Table~\ref{tab:hankel} lists the noise threshold above which each level detects the state $\rho_2$. The DWF of $\rho_2$ takes five distinct values, so by Theorem~\ref{thm:hankel} the fourth level reproduces the exact negativity threshold $1/\sqrt5$.

\begin{table}[t]
\caption{ \justifying \small Detection thresholds for $\rho_2$. 
Level $m$ of the hierarchy of Theorem~\ref{thm:hankel} detects Wigner negativity for $p$ above the listed value.}
\label{tab:hankel}
\begin{ruledtabular}
\begin{tabular}{lcc}
Criterion & Highest moment & Threshold in $p$\\
\hline
$H_1\succeq0$ (Theorem~\ref{thm:main}) & $w_3$ & 0.5806\\
$H_2\succeq0$ & $w_5$ & 0.4868\\
$H_3\succeq0$ & $w_7$ & 0.4525\\
$H_4\succeq0$ & $w_9$ & 0.4472\\
$w_2^3\le w_4$ (Theorem~\ref{thm:bound4}) & $w_4$ & 0.6224\\
Exact Wigner negativity & --- & 0.4472\\
\end{tabular}
\end{ruledtabular}
\end{table}

The hierarchy progressively closes the ququint gap (Table~\ref{tab:hankel}). Its Wigner distribution has at most five distinct values, so the fourth level reaches the exact negativity threshold. The fourth-moment witness alone detects negativity only for $p\gtrsim0.6224$, illustrating that an even-moment witness and the Hankel hierarchy have different detection properties.

\subsection{Operational quantification of magic}\label{s5}
We will now demonstrate that the same low-order Wigner moments used to certify nonstabilizerness also provide quantitative lower bounds on Wigner negativity and, consequently, the amount of magic present in a quantum state.

\subsubsection{Negativity and mana}\label{subsec:mana}

In the discrete Wigner formalism, the natural measures of magic quantify the amount of negative weight carried by the Wigner function. The sum negativity and the $\ell_1$-norm of the DWF are respectively defined as
\begin{align}
\mathcal{N}(\rho) &= \sum_{\mathbf{u}:\,W_\rho(\mathbf{u})<0} \big|W_\rho(\mathbf{u})\big|, ~\text{ and } \label{eq:sumneg}\\[2pt]
\|W_\rho\|_{1} &= \sum_{\mathbf{u}} \big|W_\rho(\mathbf{u})\big| = 1+2\,\mathcal{N}(\rho). \label{eq:l1neg}
\end{align}
The second equality follows from normalization, since the positive part of $W_\rho$ must exceed the negative part by exactly one. The logarithm of the $\ell_1$-norm is the mana, or Wigner log negativity~\cite{veitch_2014},
\begin{align}
\mathscr{M}(\rho) = \log \|W_\rho\|_{1}. \label{eq:mana}
\end{align}

Mana is additive under tensor products ($\mathscr M(\rho\otimes\sigma)=\mathscr M(\rho)+\mathscr M(\sigma)$) and nonincreasing under stabilizer operations. It vanishes precisely for states with nonnegative DWFs and provides operational bounds on magic-state distillation and the classical simulation overhead of quasiprobability-based methods~\cite{veitch_2014} which we discuss in Sec.~\ref{sec:simulation}.

Evaluating $\mathscr{M}$ generally requires reconstructing the complete DWF, which becomes increasingly demanding with system size. Here, we circumvent this requirement by deriving two complementary lower bounds on mana from low-order Wigner moments, involving $(w_2,w_3)$ and $(w_2,w_4)$, respectively. A strictly positive bound certifies nonstabilizerness while simultaneously quantifying the minimum amount of magic present in the state.

\subsubsection{Certified bound from the second and third moments}\label{subsec:bound3}

We first establish a quantitative connection between the violation of $w_2^2\leq w_3$ (Theorem~\ref{thm:main}) and the sum negativity. Recall the rescaled moments $P=Dw_2$ and $b=D^2w_3$ introduced in Eq.~\eqref{eq:rescaled}

\begin{theorem}\label{thm:bound3}
For every state $\rho$ of $n$ qudits of odd prime dimension $d$,
\begin{align}
\mathcal{N}(\rho) \;\ge\; \frac{D^2\,(w_2^2-w_3)}{1+2Dw_2+D^2w_3} \;=\; \frac{P^2-b}{1+2P+b}. \label{eq:bound3}
\end{align}
For a pure state this reads
\begin{align}
\mathcal{N}(\psi) \;\ge\; \frac{1-D^2w_3}{3+D^2w_3}. \label{eq:bound3pure}
\end{align}
\end{theorem}

The proof is provided in Methods~\ref{methods:proof_bound3}. Three observations follow from this theorem. First, the numerator of Eq.~\eqref{eq:bound3} is proportional to the violation $w_2^2-w_3$. Thus, the lower bound is strictly positive precisely when the criterion of Theorem~\ref{thm:main} is violated. Conversely, for $\mathcal{N}=0$, the bound recovers the inequality $w_2^2\leq w_3$. Second, the bound is tight when all inequalities used in the proof are saturated, which occurs when all negative DWF entries equal $-1/D$ and all positive entries are equal. The qutrit strange state provides an example: with $D^2w_3=0$, Eq.~\eqref{eq:bound3pure} yields $\mathcal{N}\geq1/3$, exactly matching its sum negativity. Finally, the bound directly provides a certified lower bound on mana,
\begin{align}
\mathscr M(\rho)\ge \mathscr M_3(\rho):=\log\!\left[1+\frac{2\,(P^2-b)}{1+2P+b}\right]. \label{eq:M3}
\end{align}

\subsubsection{Certified bound from the second and fourth moments}\label{subsec:bound4}

Theorem~\ref{thm:bound3} has a limitation: as shown in Eq.~\eqref{eq:case2}, its lower bound on the sum negativity never exceeds the purity $P\leq1$, whereas the negativity of tensor-product magic states can grow exponentially with the number of copies. To overcome this limitation, we derive a complementary bound involving the second and fourth Wigner moments.

\begin{theorem}\label{thm:bound4}
For every state $\rho$ of $n$ qudits of odd prime dimension $d$,
\begin{align}
\|W_\rho\|_1 \;\ge\; \frac{w_2^{3/2}}{w_4^{1/2}} = \sqrt{\frac{P^3}{c}}\,, \label{eq:bound4}
\end{align}
with $c=D^3w_4$, and equality holds if and only if $|W_\rho|$ is constant on its support. Equivalently,
\begin{align}
\mathscr M(\rho)\;\ge\; \mathscr M_4(\rho) := \frac12\log\frac{w_2^3}{w_4}. \label{eq:M4}
\end{align}
The quantity $\mathscr M_4$ is additive under tensor products and invariant under Clifford unitaries. For pure states $\mathscr M_4(\psi)\ge0$, with equality if and only if $\psi$ is a stabilizer state.
\end{theorem}

For a detailed proof see Methods~\ref{methods:proof_bound4}. Importantly, Eq.~\eqref{eq:bound4} holds irrespective of the signs of the DWF entries. For a nonnegative DWF, $|W|_1=1$, implying $w_2^3\leq w_4$. Thus, Theorem~\ref{thm:bound4} also yields an alternative sufficient criterion for Wigner negativity:
\begin{align}
w_2^3>w_4 \quad\Longrightarrow\quad \text{ the Wigner function of } \rho \text{ is negative}. \label{eq:crit4}
\end{align}

For the qutrit strange state $w_2^3/w_4=2$, so $\|W\|_1\ge\sqrt2\approx1.41$, against the exact value $5/3$. Similarly, for the ququint state $\ket{\psi_4}$ one finds $\|W\|_1\ge1.43$, against the exact value $1.79$.

The same argument works with any higher even moment. H\"older's inequality between $\|W\|_1$, $\|W\|_2$ and $\|W\|_{2n}$ gives
\begin{align}
\|W_\rho\|_1\ge\left(\frac{w_2^{\,2n-1}}{w_{2n}}\right)^{1/(2n-2)}, \qquad n\ge2. \label{eq:bound2n}
\end{align}
For all examples of this paper and for about $5900$ random negative states in $d=3$ and $d=5$, this family weakens as $n$ grows, and it tends to the trivial value $w_2/\max_{\mathbf u}|W_\rho(\mathbf u)|$ as $n\to\infty$. The fourth moment is thus the most useful member of the family, and we do not pursue higher orders here.

\subsubsection{Relation between the two bounds}\label{subsec:relation}

The two bounds rely on different Wigner moments and capture distinct features of the underlying distribution. Below, we compare their detection capabilities, tightness, and scaling behavior to clarify their respective advantages.

\textit{Detection.--} Theorem~\ref{thm:bound3} provides a strictly positive lower bound on the sum negativity when $w_2^2>w_3$, whereas Theorem~\ref{thm:bound4} certifies negativity when $w_2^3>w_4$. These criteria are generally inequivalent, and neither implies the other. For all examples considered in this work, the third-moment criterion is at least as robust against noise as the fourth-moment criterion. In particular, for the noisy state $\rho_2$ introduced in Sec.~\ref{s3}, the third-moment criterion detects Wigner negativity for $p>0.581$, while the fourth-moment criterion requires $p>0.622$.

\textit{Tightness.--} The two bounds are saturated under different conditions. The third-moment bound is tight when all negative Wigner entries equal $-1/D$ and all positive entries are equal, as exemplified by the qutrit strange state. In contrast, the fourth-moment bound is saturated when $|W|$ takes a single value on its support. Since the strange state has $|W|\in\{1/3,1/6\}$, the fourth-moment bound is not tight.

\begin{figure*}[t]
\centering
\includegraphics[width=0.9\textwidth]{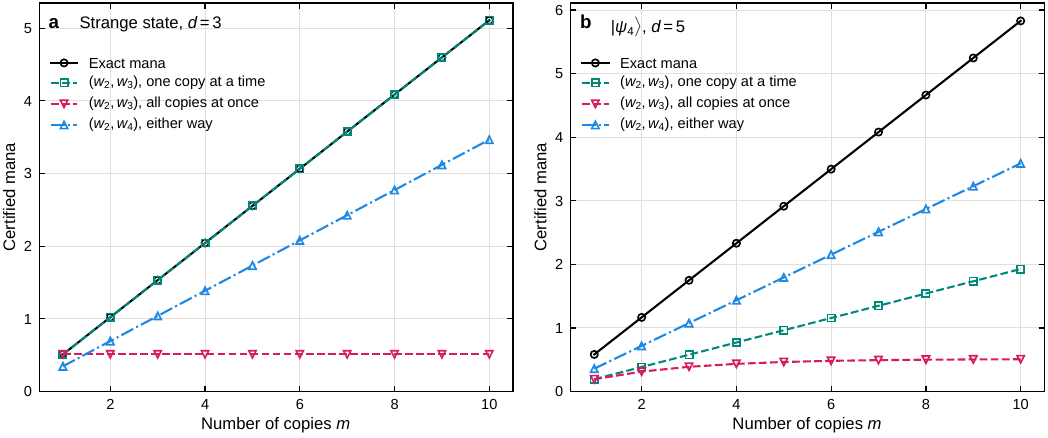}
\caption{\justifying \small Certified mana of $m$ copies of (a) the strange state and (b) $\ket{\psi_4}$. The third-moment bound of Theorem~\ref{thm:bound3} is applied either to each copy separately (teal squares) or to the global state of all copies (magenta triangles down). The fourth-moment bound of Theorem~\ref{thm:bound4} gives the same result in both cases (blue triangles up). Black circles show the exact mana. In panel (a) the per-copy third-moment bound coincides with the exact mana.}
\label{fig:copies}
\end{figure*}

\textit{Scaling with system size.--} The third-moment bound never exceeds $\mathcal N=1$ as follows from Eq.~\eqref{eq:case2}, while the fourth-moment bound is multiplicative under tensor products. Consequently, when applied directly to $m$ copies of a resource state, only the fourth-moment bound captures the exponential growth of the Wigner $\ell_1$ norm. The limitation of the first is not fundamental, however. Resource states are usually supplied as product states, and then Theorem~\ref{thm:bound3} can be applied to each copy separately and the results multiplied. Figure~\ref{fig:copies} compares the three options for the strange state and for $\ket{\psi_4}$.

\textit{Combination.--} Both bounds hold at the same time, so the best certificate is
\begin{align}
\mathcal N(\rho)\ge\max\Big\{\frac{P^2-b}{1+2P+b},\;\frac12\Big(\sqrt{\frac{P^3}{c}}-1\Big)\Big\}. \label{eq:combined}
\end{align}
The combined mana certificate is $\mathscr M(\rho)\geq\log[1+2B(\rho)]$, where $B$ is the right-hand side of Eq.~\eqref{eq:combined}.

\subsubsection{Numerical comparison}\label{subsec:numerics}

\begin{figure*}[t]
\centering
\includegraphics[width=0.92\textwidth]{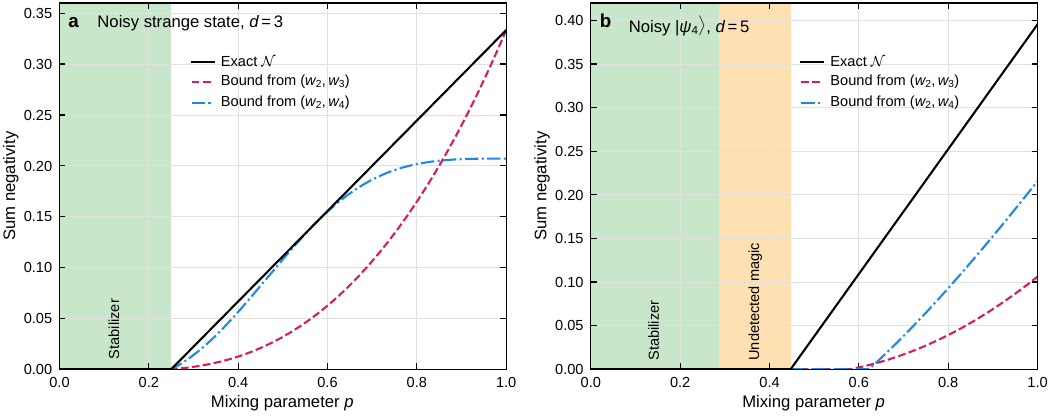}
\caption{\justifying \small Exact sum negativity and the certified lower bounds of Theorem~\ref{thm:bound3} and Theorem~\ref{thm:bound4} for (a) the noisy strange state $\rho_1$ and (b) the noisy state $\rho_2$ in $d=5$. The black line is the exact sum negativity, the magenta dashed line the bound from $(w_2,w_3)$ and the blue dash-dotted line the bound from $(w_2,w_4)$. In the green region the state is a mixture of stabilizer states. In the orange region of panel (b), labelled undetected magic, the state is magic but its Wigner function is nonnegative.}
\label{fig:bounds}
\end{figure*}

Figure~\ref{fig:bounds} compares the two bounds for the two noisy-state families introduced in Sec.~\ref{subsec:mixed}. For the noisy strange state the third-moment bound is exact at $p=1$ and the fourth-moment bound is stronger for intermediate noise. For the noisy state $\rho_2$ the fourth-moment bound is stronger throughout. Neither bound can see the orange region of Fig.~\ref{fig:bounds}(b), marked undetected magic, where the state is magic but its discrete Wigner function is nonnegative. This limitation is inherent to Wigner-negativity-based certification, as discussed in Sec.~\ref{subsec:hierarchy}.

\begin{figure*}[t]
\centering
\includegraphics[width=0.92\textwidth]{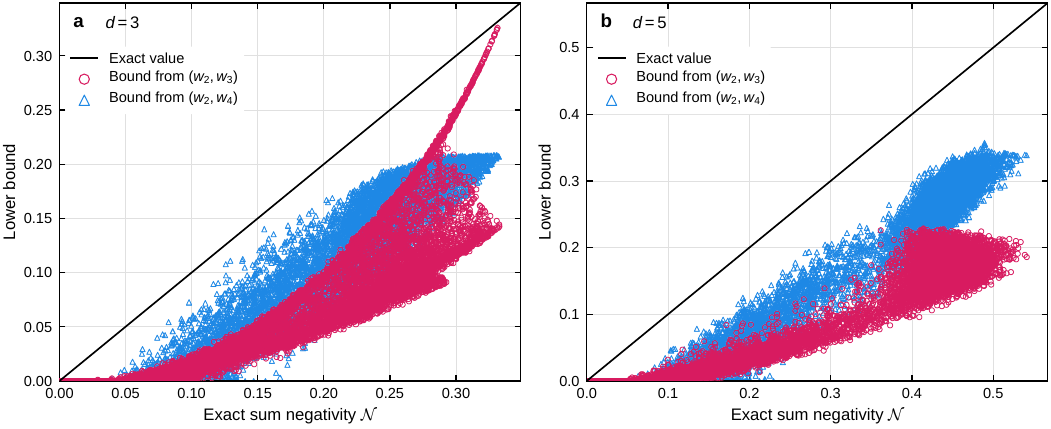}
\caption{\justifying \small Certified lower bounds against the exact sum negativity for 10000 random states in (a) $d=3$ and (b) $d=5$. Half of the states are pure and half are mixed states of random rank. No point lies above the diagonal.}
\label{fig:random}
\end{figure*}

To further assess their performance, Fig.~\ref{fig:random} compares the certified and exact sum negativities for $10\,000$ randomly generated states in dimensions $d=3$ and $d=5$, comprising equal numbers of pure and mixed states of random ranks. As expected from the theorems, all points lie on or below the diagonal. The $(w_2,w_3)$ bound exceeds the $(w_2,w_4)$ bound for only $14.6\%$ and $16.6\%$ of the negative states for $d=3$ and $d=5$, respectively. The mean ratios of the individual bounds to the exact sum negativity are $0.299$ and $0.409$ for $d=3$, and $0.246$ and $0.389$ for $d=5$, respectively. Taking the maximum of the two bounds improves these ratios to $0.419$ and $0.398$, respectively.

Finally, Figure~\ref{fig:copies} illustrates the certified mana of $m$ copies. The global third-moment bound saturates after one copy. The per-copy third-moment bound and the fourth-moment bound both grow linearly with $m$, as the exact mana does. For the strange state the per-copy bound is exact for every $m$.

\subsection{Direct estimation of discrete Wigner moments from few copies}\label{sec:multicopy}

All the results established so far rely on a few low-order moments of the DW. Their practical applicability therefore depends on how efficiently these moments can be measured. A natural route uses single-copy randomized measurements, and we study it numerically in Sec.~\ref{sec:numerics}. Here we show that a few copies of the state, measured jointly, give each moment directly, in the spirit of the multicopy estimation of nonlinear functionals~\cite{ekert_2002} and of Bell-sampling tests for stabilizer states~\cite{gross_2021}. In particular, we demonstrate that each rescaled discrete Wigner moment whose order is not a multiple of $d$ can be expressed as the expectation value of a fixed Hermitian observable with outcomes $\pm1$, enabling direct estimation without reconstructing the full DWF.

\subsubsection{Multicopy observable}\label{subsec:Rk}

Consider $k$ copies of an $n$-qudit system, with computational basis states denoted by $\ket{\mathbf x_1,\dots,\mathbf x_k}$ with $\mathbf x_j\in\Z_d^n$. We define $\mathbf s=\mathbf x_1+\dots+\mathbf x_k$ with addition performed modulo $d$.

\begin{proposition}\label{prop:Rk}
Let $k\ge2$ be an integer that is not a multiple of $d$, and let $k^{-1}$ be its inverse in $\Z_d$. Define the operator $R_k$ on $k$ copies by
\begin{align}
R_k\ket{\mathbf x_1,\dots,\mathbf x_k}=\ket{2k^{-1}\mathbf s-\mathbf x_1,\dots,2k^{-1}\mathbf s-\mathbf x_k}. \label{eq:Rk}
\end{align}
Then $R_k$ is a permutation of the computational basis, it is Hermitian and unitary with $R_k^2=\mathbb I$, and for every state $\rho$
\begin{align}
D^{\,k-1}\,w_k=\Tr\!\big[\rho^{\otimes k}R_k\big]. \label{eq:wkRk}
\end{align}
\end{proposition}

For a detailed proof see Methods~\ref{methods:proof_multicopy}. For $k=2$ Eq.~\eqref{eq:Rk} is the swap operator and Eq.~\eqref{eq:wkRk} is the familiar swap test for the purity, in agreement with Eq.~\eqref{eq:w2purity}. The operator $R_k$ acts independently on each of the $n$ qudit positions, so it is a tensor product of $n$ identical operators, each acting on $k$ qudits. The measurement circuit is given in Methods, Sec.~\ref{subsec:circuit}.

For fixed $k,d$, the Clifford transformation in Methods converts $R_k$ to a product of local parity operators. Each round prepares $k$ copies, applies this transformation across corresponding qudits, measures local parity, and multiplies the outcomes. The product is $r=\pm1$ with mean $D^{k-1}w_k$. The Clifford preprocessing has depth independent of $n$ under parallel inter-copy connectivity. Local parity readout is an additional measurement primitive, whose basis and resource assumptions are specified in Methods; it is not automatically a Clifford-only computational-basis readout.

\subsubsection{Sample complexity}\label{subsec:samples}
Each round of the multicopy protocol returns a binary outcome $r=\pm1$ whose expectation value is the rescaled Wigner moment of interest, provided that the moment order $k$ is not divisible by the local dimension $d$. We now determine the number of measurement rounds required to estimate this expectation value to a prescribed accuracy and confidence. Since the measurement outcomes are bounded
independently of the Hilbert-space dimension, the resulting sample complexity is independent of the number of qudits. The analysis follows from Hoeffding's inequality \cite{hoeffding_1963}, stated as Lemma~\ref{lem:hoeffding} in Methods~\ref{app:sample_complexity}.

Applied to the outcomes of Sec.~\ref{subsec:circuit}, whose mean is $D^{k-1}w_k$ by Eq.~\eqref{eq:meanr}, the concentration inequality, combined with Theorems~\ref{thm:main}, \ref{thm:bound3} and \ref{thm:bound4}, yields the following result.

\begin{corollary}\label{cor:samples}
\textit{(i) Measuring a single moment.} Let $k\geq2$ be an integer not divisible by $d$. The average of $r$ over $M$ independent measurement rounds estimates $D^{k-1}w_k$ to within additive error $\eta$ with confidence at least $1-\delta$ when
\begin{align}
M\geq\frac{2}{\eta^2}\ln\frac{2}{\delta}.
\label{eq:hoeffding}
\end{align}
This requires $kM$ copies of $\rho$ and is independent of the number of qudits $n$ and the local dimension $d$.

\textit{(ii) Detection.} For $d\ge5$, estimate $P$ from two-copy rounds and $b$ from three-copy rounds, $\lceil(18/\epsilon^2)\ln(4/\delta)\rceil$ rounds each, clip the estimate $\hat P$ to $[0,1]$, and set $\hat G=\hat P^2-\hat b$. This uses
\begin{equation}
N=5\left\lceil\frac{18}{\epsilon^2}\ln\frac4\delta\right\rceil
\label{eq:Mtot}
\end{equation}
copies of $\rho$, independently of the number of qudits, and
\begin{equation}
\Pr(|\hat G-G|>\epsilon)\leq\delta.
\label{eq:witness_confidence_main}
\end{equation}
Declaring the state magic whenever $\hat G>\epsilon$ is therefore incorrect with probability at most $\delta$, and a state with $G>2\epsilon$ is detected with probability at least $1-\delta$.

\textit{(iii) Quantifier from $(w_2,w_3)$.} For $d\ge5$, estimate $P$ and $b$ from $M=\lceil(2/\eta^2)\ln(4/\delta)\rceil$ rounds each, which uses $5M$ copies of $\rho$, and set $b_+=\hat b+\eta$ and $\underline{P}=\max\{0,\hat P-\eta,-b_+\}$. Then, with probability at least $1-\delta$,
\begin{equation}
\mathcal N(\rho)\ge\frac{\underline{P}^2-b_+}{1+2\underline{P}+b_+}.
\label{eq:cert_bound3}
\end{equation}

\textit{(iv) Quantifier from $(w_2,w_4)$.} For every odd prime $d$, estimate $P$ and $c$ from $M=\lceil(2/\eta^2)\ln(4/\delta)\rceil$ rounds each, which uses $6M$ copies of $\rho$, and set $c_+=\hat c+\eta$. If $c_+>0$, then, with probability at least $1-\delta$,
\begin{equation}
\mathcal N(\rho)\ge\frac12\left[\sqrt{\frac{\max\{0,\hat P-\eta\}^3}{c_+}}-1\right],
\label{eq:cert_bound4}
\end{equation}
or equivalently $\|W_\rho\|_1\ge\sqrt{\max\{0,\hat P-\eta\}^3/c_+}$.
\end{corollary}

The proof is provided in Methods~\ref{methods:proof_sampling}. The rescaled moments $P$, $b$ and $c$ defined in Eq.~\eqref{eq:rescaled} are exactly the quantities measured by this protocol for $k=2,3,4$. These moments directly determine the magic-detection and quantification criteria established in Sec.~\ref{s5}. To see how the corollary translates into an operational certificate, consider the criterion $P^2-b>0$.
The estimates of $P$ and $b$ are each accurate to $\epsilon/3$ with failure probability $\delta/2$. Clipping $\hat P$ to the interval $[0,1]$ in which $P$ must lie can only reduce its error, and then $|\hat P+P|\le2$, so the error of $\hat P$ enters the witness at most twice and that of $\hat b$ once. By the union bound the witness is therefore accurate to $\epsilon$ with probability at least $1-\delta$, which gives Eqs.~\eqref{eq:Mtot} and \eqref{eq:witness_confidence_main}.

For a pure magic state, the violation is $1-D^2w_3$. Whenever this violation remains bounded away from zero, magic can be certified with a number of copies independent of the system size.

The detailed error-propagation analysis, including general individual accuracies, failure probabilities, and the union bound, is provided in Methods~\ref{methods:proof_sampling}.

The certified bounds of Theorems~\ref{thm:bound3} and \ref{thm:bound4} are handled in parts (iii) and (iv), by inserting the lower or upper endpoints of the corresponding confidence intervals where they make the bounds smaller, namely the upper endpoints for $b$ and $c$ and the lower endpoint for $P$. Since $P+b\ge0$ by Eq.~\eqref{eq:rescaledbounds}, in part (iii) the lower endpoint for $P$ is replaced by $-b_+$ whenever it lies below it, which keeps the $(w_2,w_3)$ quantifier valid. The same holds for any one-sided confidence intervals $P\ge P_-$, $b\le b_+$ and $c\le c_+$ that are valid together with probability at least $1-\delta$, such as those obtained from the classical-shadow estimates of Sec.~\ref{sec:numerics}, with $\hat P-\eta$, $\hat b+\eta$ and $\hat c+\eta$ replaced by $P_-$, $b_+$ and $c_+$. Since part (iv) uses only $P$ and $c$, which have binary multicopy observables for every odd prime $d$, the $(w_2,w_4)$ quantifier is available for qutrits as well. In contrast, estimating global nonlinear functionals from randomized single-copy measurements generally requires a number of measurements that grows exponentially
with $n$~\cite{huang_2020,elben_2020}. For strongly mixed states, the purity $P$ itself can become exponentially small in $n$, requiring correspondingly higher precision in the witness estimation.

\begin{figure}[t]
\centering
\includegraphics[width=\linewidth]{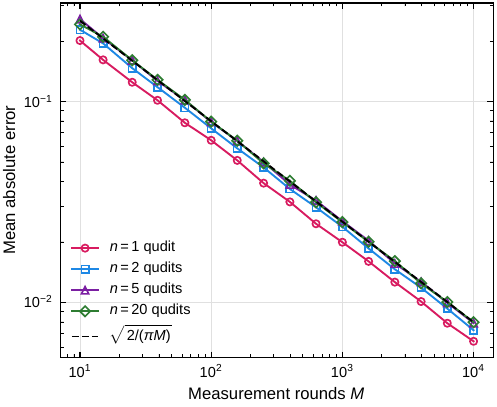}
\caption{\justifying \small Mean absolute error of the estimate of $D^2w_3$ from $M$ rounds of the three-copy protocol of Sec.~\ref{subsec:circuit} for $\ket{\psi_4}^{\otimes n}$ with $d=5$, averaged over 4000 repetitions, for $n=1$ (magenta circles), $n=2$ (blue squares), $n=5$ (purple triangles) and $n=20$ (green diamonds). The black dashed line is the large-$M$ value for an unbiased $\pm1$ variable of zero mean, which is the worst case.}
\label{fig:sampling}
\end{figure}

Figure~\ref{fig:sampling} shows a simulation of the three-copy protocol for the product state $\ket{\psi_4}^{\otimes n}$, for which $D^2w_3=0.616^{\,n}$. The outcomes $r=\pm1$ are drawn with the exact mean of Eq.~\eqref{eq:meanr}. The estimation error exhibits nearly identical scaling
with the number of measurement rounds for systems ranging from one to twenty qudits, consistent with the dimension-independent sample-complexity bound predicted by Corollary~\ref{cor:samples}.

\begin{remark}\label{rem:qutrit}
When $k$ is a multiple of $d$, the condition in Eq.~\eqref{eq:sumA} admits $D$ solutions
for $\mathbf s=\mathbf 0$ and none otherwise. In this case, the rescaled moment can be
expressed as $D^{k-1}w_k=D\,\Tr[\rho^{\otimes k}O_k]$ with an observable $O_k$ of eigenvalues
$\pm1$ and $0$. The variance of the resulting estimate is at most $D^2\,\Tr[\rho^{\otimes k}O_k^2]$ per round.

The operator $\sum_{\mathbf u}A_{\mathbf u}^{\otimes k}$ commutes with $U^{\otimes k}$ for every
Clifford unitary $U$, so this quantity is the same for all stabilizer states, and for them it grows linearly with $D$. For other states, however, it can be considerably smaller, as illustrated
by the tensor-product strange states considered in Sec.~\ref{sec:numerics}.

This distinction is particularly relevant for estimating the third Wigner moment in qutrit systems, where $d=k=3$. For $d=3$, we therefore recommend the pair $(w_2,w_4)$, for which
Proposition~\ref{prop:Rk} holds and Theorem~\ref{thm:bound4} provides both the detection criterion and the bound. For the noisy strange state, the criterion $w_2^3>w_4$ detects negativity for all
$p>1/4$, exactly as $w_2^2>w_3$ does.
\end{remark}


\subsection{Numerical simulation: estimating the moments in practice}\label{sec:numerics}

The multicopy protocol of Sec.~\ref{sec:multicopy} provides direct access to discrete Wigner moments through joint measurements on a few copies of the state. Many current platforms, however, can only measure one copy at a time. For them the standard tool for nonlinear functions of a state is the randomized measurement~\cite{vanenk_2012,brydges_2019,elben_2023}, and its most flexible form is the classical shadow~\cite{huang_2020}. In this section, we develop unbiased classical-shadow estimators of discrete Wigner moments, investigate their performance for the noisy strange state, and compare their measurement costs with those of conventional quantum state tomography. The classical-shadow construction is given in Methods~\ref{subsec:shadows}.

\subsubsection{A single qutrit}\label{subsec:qutritnum}

\begin{figure*}[t]
\centering
\includegraphics[width=\textwidth]{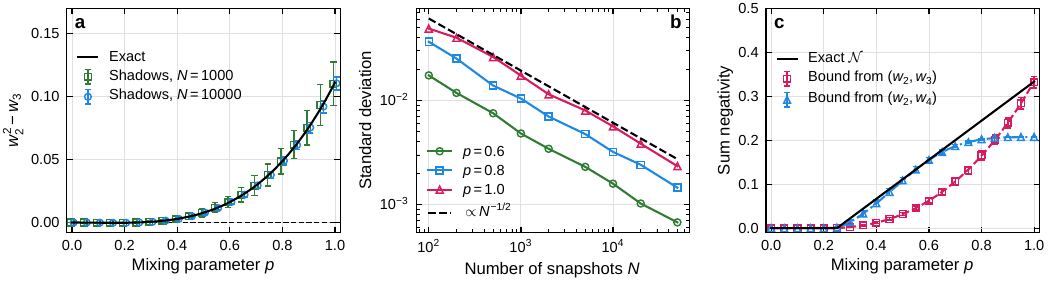}
\caption{\justifying \small Classical-shadow estimation for the noisy strange state $\rho_1$. (a) Estimated violation $w_2^2-w_3$ for $N=10^3$ (green squares) and $N=10^4$ (blue circles) snapshots, mean and standard deviation over $200$ repetitions, with the exact curve (black line). (b) Standard deviation of the estimated violation as a function of the number of snapshots for $p=0.6$ (green circles), $p=0.8$ (blue squares) and $p=1$ (magenta triangles). The black dashed line is proportional to $N^{-1/2}$. (c) Certified lower bounds on the sum negativity from Theorems~\ref{thm:bound3} and \ref{thm:bound4}, estimated from $N=10^4$ snapshots, the $(w_2,w_3)$ bound as magenta squares and the $(w_2,w_4)$ bound as blue triangles. The magenta dashed and blue dash-dotted lines are the exact bounds, and the black line is the exact sum negativity.}
\label{fig:shadowqutrit}
\end{figure*}

We first estimate the moments of the noisy strange state $\rho_1=p\ket \psi_1\!\bra \psi_1+(1-p)\mathbb I_3/3$ of Sec.~\ref{s3}. Figure~\ref{fig:shadowqutrit}(a) shows the estimated violation $w_2^2-w_3$ over $200$ repetitions for $N=10^3$ and $N=10^4$ snapshots. The estimates follow the exact curve $w_2^2-w_3=p^2(p+2)(4p-1)/81$ at every $p$, with no visible bias. For $N=10^4$, the estimated violation exceeds three standard deviations for all $p\geq0.35$, demonstrating reliable magic detection even in the presence of substantial noise.

Figure~\ref{fig:shadowqutrit}(b) shows the estimator's standard deviation on the number of snapshots $N$ for three representative magic states. The standard deviation decreases as $N^{-1/2}$, consistent with the expected scaling of a nondegenerate U-statistic. Interestingly, for the maximally mixed state ($p=0$), which is not shown in the figure, a faster scaling of $N^{-1}$ is observed. This behavior arises because the leading statistical fluctuations of a U-statistic are determined by the snapshot-to-snapshot variation of the overlap $\sum_{\mathbf u}\hat W_i(\mathbf u)W_\rho(\mathbf u)$.  
A flat Wigner function makes this overlap the same for every snapshot, so only the next-order term survives.

Figure~\ref{fig:shadowqutrit}(c) presents the lower bounds on the sum negativity obtained from the estimated moments using Theorems~\ref{thm:bound3} and \ref{thm:bound4}, with $N=10^4$ snapshots. 
On average the estimated certificates reproduce the exact bounds. Single estimates fluctuate around them, so a rigorous certificate uses the lower end of a confidence interval, as described in Sec.~\ref{subsec:samples}. This behaviour should be contrasted with a naive plug-in estimate of the negativity itself, $\sum_{\mathbf u}|\overline{W}(\mathbf u)|$ with $\overline W$ the averaged snapshot. The absolute value turns statistical noise around zero into apparent negativity, so this estimate is biased upward. With $10^3$ snapshots it reports $\|W\|_1\approx1.002$ at $p=0.2$, where the Wigner function is nonnegative. The moment-based certificates avoid this problem because they are built from unbiased estimates of smooth quantities.

\subsubsection{Comparison with state tomography}\label{subsec:tomography}

The obvious alternative to estimating the moments is to reconstruct the state and compute the moments from the reconstruction. To compare these approaches, we consider a register of $n$ noisy strange states, $\rho=\rho_1^{\otimes n}$ with $p=0.9$, for $n=1,2,3,4$, corresponding to Hilbert-space dimensions $D=3,9,27,81$. Such a register is what a magic state factory delivers, and its rescaled violation $V=D^2(w_2^2-w_3)=P^2-b$ is of order unity for every $n$. We estimate $V$ in four ways and count the total number of copies $N$ used by each.
\begin{itemize}
\item Global shadows, with the estimators of Eqs.~\eqref{eq:hatw2} and \eqref{eq:hatw3}.
\item Local shadows, with the same estimators.
\item Standard state tomography. Each qutrit is measured in one of its four mutually unbiased bases~\cite{wootters_1989}, all $4^n$ settings are used equally often, and the state is reconstructed by linear inversion. The reconstruction is projected onto the set of density matrices~\cite{smolin_2012}, and $V$ is computed from its DWF.
\item The multicopy protocol of Sec.~\ref{sec:multicopy}, with two-copy rounds for $P$ and three-copy rounds for $b$. For qutrits the three-copy observable has the eigenvalues $\pm1$ and $0$ (Remark~\ref{rem:qutrit}). Its variance, and hence the error, is computed exactly for this state, with the copies split optimally between the two kinds of rounds.
\end{itemize}

\begin{figure*}[t]
\centering
\includegraphics[width=0.9\textwidth]{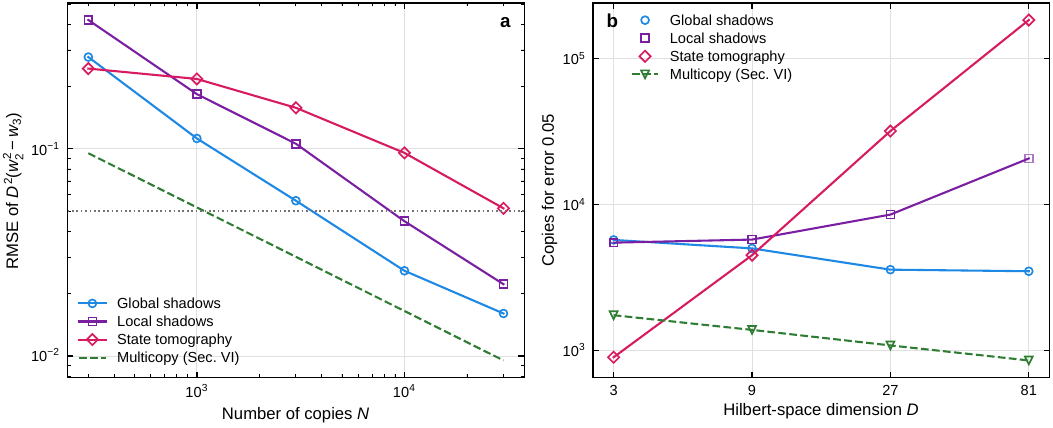}
\caption{\justifying \small Cost of estimating the rescaled violation $V=D^2(w_2^2-w_3)$ for a register of $n$ noisy strange states with $p=0.9$. Blue circles show global shadows, purple squares local shadows, magenta diamonds state tomography, and the green dashed line (triangles) the multicopy protocol of Sec.~\ref{sec:multicopy}. (a) Root-mean-square error as a function of the total number of copies for three qutrits ($D=27$), over the range of $N$ simulated for all methods. The grey dotted line marks the target error $0.05$. (b) Number of copies needed to reach the error $0.05$, as a function of the dimension $D=3^n$, interpolated from the simulated data. The lighter square for local shadows at $D=81$ is extrapolated from the last two simulated points. The multicopy values follow from the exact variances of the outcomes.}
\label{fig:tomocost}
\end{figure*}

Figure~\ref{fig:tomocost}(a) shows the root-mean-square error of each estimate for three qutrits. The three moment-based estimates converge essentially as $N^{-1/2}$. Tomography converges more slowly, because its error is dominated by a systematic bias rather than by fluctuations. The reconstruction has $D^2$ noisy parameters, and the projection onto physical states pulls the reconstructed discrete Wigner function towards zero, so the violation is underestimated until the noise per parameter becomes small. The bias shrinks only as the number of copies per parameter grows.

Figure~\ref{fig:tomocost}(b) shows the number of copies needed to reach an error of $0.05$, a precision that resolves $V$ for every $n$ considered. For a single qutrit tomography is the cheapest method, which is not surprising, since a qutrit has only eight real parameters. From two qutrits onward the picture changes. The cost of tomography grows from about $10^3$ copies at $D=3$ to about $2\times10^5$ at $D=81$, roughly as $D^{1.6}$. The cost of global shadows stays between $3.5\times10^3$ and $6\times10^3$ over the whole range. Local shadows grow slowly, and at $D=81$ they need about nine times fewer copies than tomography. The multicopy protocol needs between $0.9\times10^3$ and $1.8\times10^3$ copies, and its cost even decreases with $n$ for this state, because the variance of the three-copy observable shrinks for a register of strange states.

These numbers must be read with the right caveats. Global random unitaries generally require deeper circuits as the number of qudits increases, whereas local random unitaries are considerably easier to implement. Moreover, the variance of the global-shadow estimator contains a contribution of order $D^2/N^2$, which is negligible for the dimensions considered here but can become significant for larger systems, requiring $N\gtrsim D$ to suppress this contribution. For qutrits the three-copy observable is not a $\pm1$ variable, and for some states, stabilizer states among them, its variance grows linearly with $D$ (Remark~\ref{rem:qutrit}). In contrast, for odd prime dimensions $d\geq5$, all Wigner moments up to fourth order admit binary-outcome multicopy measurements, for which Corollary~\ref{cor:samples} guarantees dimension-independent sample complexity at fixed additive precision, whereas a direct estimate of every phase-space value of the Wigner function needs a number of copies that grows as $D^2$ (see Supplementary Information, Comparison with direct Wigner tomography). Furthermore, the estimators involve classical post-processing over $D^2$ phase-space points, introducing additional computational costs. Despite these limitations, our numerical results demonstrate that estimating low-order Wigner moments can require substantially fewer copies than full state tomography for magic certification, with the advantage increasing with system size over the range investigated.


\subsection{Application: certified cost of classical simulation}\label{sec:simulation}
Stabilizer circuits with nonnegative DWF representations admit efficient classical simulation~\cite{gottesman_1998,mari_2012,veitch_2012}, whereas magic states provide the additional resources required for universal quantum computation~\cite{Bravyi_2005}. DWF negativity thus connects these resources to the cost of classical simulation: quasiprobability methods incur a sampling overhead that grows with the negativity~\cite{pashayan_2015}. Our resource bounds constrain this overhead. In particular, for Clifford evolution and computational-basis measurement with $m$ copies of a resource state $\rho_{\rm m}$ and stabilizer states on the remaining inputs, the signed estimator defined in Methods, Sec.~\ref{methods:quasi-simulation}, admits the sufficient Hoeffding sampling prescription
\begin{equation}
s_{\rm H}(\epsilon,\delta)
=\frac{2}{\epsilon^2}\|W_{\rho_{\rm m}}\|_1^{2m}
\ln\frac{2}{\delta},
\label{eq:prescription_main}
\end{equation}
where $\epsilon$ is the target additive accuracy and $1-\delta$ is the confidence level. 

\subsubsection{Certified sample count}\label{subsec:certcost}
Evaluating Eq.~\eqref{eq:pwbm} generally requires knowledge of the complete Wigner function of $\rho_{\rm m}$. The moment-based bounds derived in Sec.~\ref{s5} provide a way to certify the corresponding sampling overhead without reconstructing the distribution.

\begin{corollary}\label{cor:cost}
Let $L(\rho_{\rm m})$ be the larger of the two certified values of $\|W_{\rho_{\rm m}}\|_1$, i.e.,
\begin{align}
L\equiv\max\Big\{1+\frac{2\,(P^2-b)}{1+2P+b},\ \sqrt{\frac{P^3}{c}}\Big\}, \label{eq:Lcert}
\end{align}
computed from the moments of a single copy of $\rho_{\rm m}$. Then the number of samples in Eq.~\eqref{eq:pwbm} obeys
\begin{align}
s_{\rm H}(\epsilon,\delta)\;\ge\;\frac{2}{\epsilon^{2}}\,L^{2m}\,\ln\frac{2}{\delta}. \label{eq:costbound}
\end{align}
\end{corollary}

The statement follows from Theorems~\ref{thm:bound3} and \ref{thm:bound4} and applies to both pure and mixed states. Noisy magic states are exactly what a real device supplies, and for them Theorem~\ref{thm:faithful} no longer applies, but Corollary~\ref{cor:cost} still does. Second, the low-order moments of a single resource state determine a certified exponential growth rate for arbitrarily many copies. Thus, estimating only three moments from a few copies of the resource is sufficient to certify the scaling of the sample-count prescription. Importantly, this result concerns the sampling overhead associated with the specific quasiprobability estimator and its concentration bound, rather than a fundamental lower bound on the complexity of arbitrary classical simulation algorithms.

\begin{figure}[t]
\centering
\includegraphics[width=0.9\linewidth]{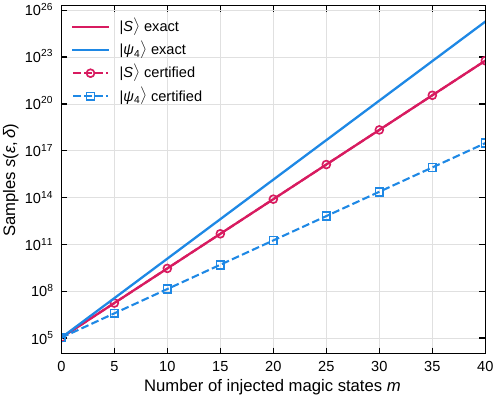}
\caption{\justifying \small Number of samples $s(\epsilon,\delta)$ needed by the quasiprobability estimator of Eq.~\eqref{eq:estimator} for $m$ injected magic states, with $\epsilon=\delta=0.01$. Solid lines use the exact $\|W\|_1$ and dashed lines with open symbols the certified value of Corollary~\ref{cor:cost}, for the strange state (magenta, circles) and $\ket{\psi_4}$ (blue, squares). For the strange state the two lines coincide.}
\label{fig:cost}
\end{figure}

We illustrate Corollary~\ref{cor:cost} using the two magic states introduced in Sec.~\ref{s3}: the qutrit strange state and the state $\ket{\psi_4}$ in $d=5$. For the strange state the bound of Theorem~\ref{thm:bound3} is exact, so the certified value equals the true one, $L=\|W\|_1=5/3$. For $\ket{\psi_4}$ the larger certificate comes from Theorem~\ref{thm:bound4}, $L=\sqrt{125/61}\approx1.43$, against the exact value $\|W\|_1\approx1.79$. Figure~\ref{fig:cost} shows the resulting sample counts as a function of the number $m$ of injected copies, for $\epsilon=\delta=0.01$. Both the exact and the certified counts grow exponentially, and the moments of a single copy certify the full rate for the strange state and $61\%$ of it for $\ket{\psi_4}$. For forty injected copies of $\ket{\psi_4}$ this is a certified count of more than $10^{17}$ samples, obtained from two-copy and four-copy measurements on a single qudit.

\section{Discussion} \label{s6}
Low-order moments of the DWF provide direct access to the operational detection and quantification of quantum magic without requiring full state reconstruction of individual phase-space values or optimization over the stabilizer polytope. For pure states of any number of odd-prime-dimensional qudits, a third moment inequality gives a necessary and sufficient criterion for nonstabilizerness. For mixed states, the elementary witness is sufficient but not necessary for Wigner negativity, and the Hankel hierarchy becomes complete for this particular property at a finite order. Beyond detection, we obtained complementary lower bounds on the sum negativity and mana from low-order moments, which further certify the sampling overhead of quasiprobability-based classical simulation with magic-state inputs.

Importantly, our framework provides an experimentally accessible route to magic certification through multicopy measurements implemented using Clifford circuits. For moments whose order is not a multiple of $d$, which includes all moments used here when $d\ge5$ and $w_2$ and $w_4$ when $d=3$, the number of measurement rounds required to achieve a fixed additive precision is independent of the system size, in contrast with the exponentially growing cost of full state reconstruction. We further demonstrated the applicability of our approach to representative qutrit and ququint states, including noisy magic states, and investigated an alternative implementation using single-copy randomized measurements. Our results establish low-order Wigner moments as operational probes of nonstabilizer resources, connecting experimentally accessible measurements to magic quantification and the cost of classical simulation. More broadly, this approach provides a promising direction for resource certification in higher-dimensional quantum systems and scalable fault-tolerant quantum information processing.

Several open avenues for future research emerge from this work. The first concerns the information that the present bound leaves aside. Combining higher-order moments may yield tighter bounds and extend the framework to other magic measures, such as robustness of magic, and to the magic of channels. Secondly, extending the approach to qubits, where no Wigner representation is covariant under the full Clifford group, may require alternative phase-space representations or stabilizer R\'enyi entropies. Finally, on the experimental side, trapped-ion and superconducting qudit processors~\cite{ringbauer_2022,morvan_2021}, which support the operations required for multicopy measurements and classical-shadow protocols, provide promising platforms for experimental demonstrations of our magic-certification framework.

\section{Methods}\label{sec:methods}
\subsection{Proofs of the detection and quantification results}\label{methods:proofs}
\subsubsection{Proof of Lemma~\ref{lem:basic}}\label{methods:proof_basic}
We prove the properties one by one.

\textit{(i) Bound on the discrete Wigner function.} The point operator at the origin is the parity operator, which reflects every computational basis state,
\begin{align}
A_{\mathbf 0}\ket{\mathbf x}=\ket{-\mathbf x}. \label{eq:parity}
\end{align}
Applying the reflection twice gives the identity, so $A_{\mathbf 0}$ has eigenvalues $\pm1$. Every other point operator is unitarily equivalent to $A_{\mathbf 0}$ by Eq.~\eqref{eq:A_u} and has the same eigenvalues. Hence
\begin{align}
-\mathbb I\le A_{\mathbf u}\le \mathbb I
\quad\Longrightarrow\quad
-1\le \Tr\big[A_{\mathbf u}\,\rho\big]\le 1 .
\end{align}
Dividing by $D$ gives Eq.~\eqref{eq:Wbound}.
 
\textit{(ii) The second moment.} The point operators are orthogonal~\cite{gross_2006},
\begin{align}
\Tr\big[A_{\mathbf u}A_{\mathbf v}\big]=D\,\delta_{\mathbf u,\mathbf v}. \label{eq:ortho}
\end{align}
There are $D^2$ of them, so they form a basis of the operator space. Expanding $\rho$ in this basis and using Eq.~\eqref{eq:ortho} to read off the coefficients, we find that the coefficients are exactly the values of the DWF,
\begin{align}
\rho=\sum_{\mathbf u}W_\rho(\mathbf u)\,A_{\mathbf u}. \label{eq:expansion}
\end{align}
Inserting Eq.~\eqref{eq:expansion} twice into the purity and using Eq.~\eqref{eq:ortho} once more,
\begin{align}
\Tr\big[\rho^2\big]
&=\sum_{\mathbf u,\mathbf v}W_\rho(\mathbf u)\,W_\rho(\mathbf v)\,\Tr\big[A_{\mathbf u}A_{\mathbf v}\big]\nonumber\\
&=D\sum_{\mathbf u}W_\rho(\mathbf u)^2 = D\,w_2,
\end{align}
which is Eq.~\eqref{eq:w2purity}.
 
\textit{(iii) Higher moments.} We keep two factors of $W_\rho(\mathbf u)$ and bound the remaining $k-2$ factors with Eq.~\eqref{eq:Wbound},
\begin{align}
|w_k|
&\le \sum_{\mathbf u}\big|W_\rho(\mathbf u)\big|^{k-2}\,W_\rho(\mathbf u)^2\nonumber\\
&\le \frac{1}{D^{\,k-2}}\sum_{\mathbf u}W_\rho(\mathbf u)^2
=\frac{w_2}{D^{\,k-2}} ,
\end{align}
which is Eq.~\eqref{eq:wkbound}.

\subsubsection{Proof of Theorem~\ref{thm:faithful}}\label{methods:proof_pure}
Throughout the proof, we denote $W_\psi$ by $W$. Now, for a pure state Eq.~\eqref{eq:w2purity} gives
\begin{align}
w_2=\frac{1}{D}. \label{eq:w2pure}
\end{align}
Each value of the DWF is at most $1/D$ by Eq.~\eqref{eq:Wbound}, and each square $W(\mathbf u)^2$ is nonnegative. Therefore
\begin{align}
w_3
&=\sum_{\mathbf u}W(\mathbf u)\,W(\mathbf u)^2\nonumber\\
&\le\frac{1}{D}\sum_{\mathbf u}W(\mathbf u)^2
=\frac{1}{D}\,w_2
=\frac{1}{D^2}, \label{eq:w3chain}
\end{align}
which is Eq.~\eqref{eq:w3pure}.
 
Next, the gap in Eq.~\eqref{eq:w3chain} can be written as a sum over phase space,
\begin{align}
\frac{1}{D^2}-w_3=\sum_{\mathbf u}W(\mathbf u)^2\Big[\frac{1}{D}-W(\mathbf u)\Big]. \label{eq:gap}
\end{align}
Every term on the right is nonnegative. The sum vanishes only if every term vanishes, that is, only if at each point
\begin{align}
W(\mathbf u)=0 \qquad\text{or}\qquad W(\mathbf u)=\frac{1}{D}. \label{eq:twovalues}
\end{align}
Such a DWF has no negative values. By the discrete Hudson theorem~\cite{gross_2006}, a pure state with a nonnegative DWF is a stabilizer state.
 
The DWF of a pure stabilizer state equals $1/D$ on exactly $D$ points of phase space and vanishes elsewhere~\cite{gross_2006}. Hence
\begin{align}
w_3=D\cdot\frac{1}{D^3}=\frac{1}{D^2}.
\end{align}
Combining Eq.~\eqref{eq:w2pure}
\begin{align}
w_2^2-w_3=\frac{1-D^2w_3}{D^2}, \label{eq:violationpure}
\end{align}
which is negative exactly when $D^2w_3>1$, that is, exactly when $\ket\psi$ is not a stabilizer state.

\subsubsection{Proof of Theorem~\ref{thm:main}}\label{methods:proof_mixed}
Let us assume that the DWF, $W_\rho(\textbf{u}) \equiv W$ is non-negative. Since $W\ge0$, we may take square roots, and $W^2 = W^{\frac{1}{2}}\cdot W^{\frac{3}{2}}$. We apply the Cauchy-Schwarz inequality,
\begin{align}
    w_2^2 &= \left( \sum_{\mathbf{u}} W^{\frac{1}{2}}\cdot W^{\frac{3}{2}} \right)^2 \nonumber\\
    & \le \left( \sum_{\mathbf{u}} W \right) \left( \sum_{\mathbf{u}} W^3 \right) = w_3.
\end{align}
This completes the proof. 
\subsubsection{Proof of lemma \ref{lemma:mixed}}\label{methods:proof_stabilizer}
\begingroup
\renewcommand{\proof}{\par\noindent\textit{}\enspace}

\proof Let $\mathrm{STAB}$ denote the set of stabilizer states in an odd prime or prime-power dimension. Now, consider a mixed stabilizer state $\rho=\sum_j p_j |S_j\rangle\langle S_j| \in \mathrm{STAB}, \text{ with } p_j\geq0, \text{ and }\sum_jp_j=1$.  By linearity of the DWF representation,
\begin{align}
        W_{\rho}(\textbf{u})&=\frac{1}{d}\Tr[A_\textbf{u}\rho] \nonumber\\
        &= \frac{1}{d}\Tr\left[A_\textbf{u}\sum_{j}p_j|S_j\rangle\langle S_j|\right] \nonumber\\
        &= \sum_{j}p_j\frac{1}{d}\Tr\left[A_\textbf{u}|S_j\rangle\langle S_j|\right] \nonumber\\
        &= \sum_{j}p_jW_{|S_j\rangle\langle S_j|}(\textbf{u})
\end{align}
Since every $|S_j\rangle\in\mathrm{STAB}$ has a non-negative DWF,  $W_{|S_j\rangle\langle S_j|}(\mathbf{u})\geq0, ~~ \forall\,j,\mathbf{u}$, and $p_j\geq0$, we immediately obtain 
\begin{equation} W_{\rho}(\mathbf{u}) =\sum_jp_jW_{\rho_j}(\mathbf{u}) \geq0, \qquad \forall\,\mathbf{u}\in\mathbb{Z}_d\times\mathbb{Z}_d. \nonumber
\end{equation} Hence, any mixed stabilizer states also possesses a non-negative DWF. However, its converse is not always true~\cite{gross_2006}. Combined with Theorem~\ref{thm:main}, this shows that every mixed stabilizer state satisfies $w_2^2\le w_3$, so a violation of this inequality certifies magic also for mixed states.
\endgroup
\subsubsection{Proof of Theorem~\ref{thm:hankel}}\label{methods:proof_hierarchy}
\begingroup
\renewcommand{\proof}{\par\noindent\textit{Proof.}\enspace}

For any real vector $\mathbf c=(c_0,\dots,c_m)$ let $f(x)=\sum_{i}c_ix^i$. Then
\begin{align}
\mathbf c^{T}H_m\,\mathbf c 
&=\sum_{i,j}c_ic_j\sum_{\mathbf u}W(\mathbf u)^{i+j+1} \nonumber \\ 
&=\sum_{\mathbf u}W(\mathbf u)\,f\big(W(\mathbf u)\big)^2. \label{eq:quadform}
\end{align}
If $W\ge0$ every term is nonnegative, which proves the first claim. For the converse let $t_1,\dots,t_r$ be the distinct values of $W$, with $t_1<0$, and choose $f(x)=\prod_{l=2}^{r}(x-t_l)$, a polynomial of degree $r-1$. Then $f$ vanishes at every value of $W$ except $t_1$, and Eq.~\eqref{eq:quadform} becomes $\mu_1\,t_1\,f(t_1)^2<0$, where $\mu_1\ge1$ is the number of points at which $W=t_1$.
\endgroup
\subsubsection{Proof of Theorem~\ref{thm:bound3}}\label{methods:proof_bound3}
 
Theorem~\ref{thm:main} follows from applying the Cauchy-Schwarz inequality to a probability distribution. Although a Wigner function with negative values is not a probability distribution, we apply the Cauchy-Schwarz inequality to the positive part only and bound the contributions from the negative part.

With this aim, we first split the DWF as $W=W_+-W_-$, where $W_+$ and $W_-$ are nonnegative and are never both nonzero at the same point. The normalization condition gives,
\begin{align}
\sum_{\mathbf u}W_- = \mathcal N, \qquad \sum_{\mathbf u}W_+ = 1+\mathcal N. \label{eq:parts}
\end{align}
Since the supports are disjoint, the moments split as
\begin{align}
w_2=\sum_{\mathbf u}W_+^2+\sum_{\mathbf u}W_-^2, \qquad
w_3=\sum_{\mathbf u}W_+^3-\sum_{\mathbf u}W_-^3. \label{eq:splitmoments}
\end{align}

Since every value of $W_-$ lies between $0$ and $1/D$ as of Lemma~\ref{lem:basic}, hence
\begin{align}
\sum_{\mathbf u}W_-^2\le \frac{\mathcal N}{D}, \qquad \sum_{\mathbf u}W_-^3\le \frac{\mathcal N}{D^2}. \label{eq:negparts}
\end{align}
Inserting this in Eq.~\eqref{eq:splitmoments} and multiplying by $D$ and $D^2$, respectively,
\begin{align}
D\sum_{\mathbf u}W_+^2 \ge P-\mathcal N, \qquad D^2\sum_{\mathbf u}W_+^3\le b+\mathcal N. \label{eq:posparts}
\end{align}

Further, applying the Cauchy-Schwarz inequality to the nonnegative distribution $W_+$ yields (total weight $1+\mathcal N$),
\begin{align}
\Big(\sum_{\mathbf u}W_+^2\Big)^{2}\le \Big(\sum_{\mathbf u}W_+\Big)\Big(\sum_{\mathbf u}W_+^3\Big)=(1+\mathcal N)\sum_{\mathbf u}W_+^3. \label{eq:CSplus}
\end{align}

Now, suppose first that $\mathcal N\le P$, so that the first bound in Eq.~\eqref{eq:posparts} is nonnegative and may be squared. Multiplying Eq.~\eqref{eq:CSplus} by $D^2$ and using Eq.~\eqref{eq:posparts} on both sides,
\begin{align}
(P-\mathcal N)^2 \le (1+\mathcal N)(b+\mathcal N). \label{eq:quad}
\end{align}
After simplification, we obtain,
\begin{align}
P^2-b\le \mathcal N\,(1+2P+b). \label{eq:linear}
\end{align}
Lemma~\ref{lem:basic} gives $b\ge-P$, so the bracket is positive and Eq.~\eqref{eq:bound3} follows.

Now, if $\mathcal N>P$, the bound holds trivially, because its right-hand side never exceeds $P$,
\begin{align}
P-\frac{P^2-b}{1+2P+b}=\frac{(1+P)(P+b)}{1+2P+b}\ge0. \label{eq:case2}
\end{align}
Setting $P=1$ gives Eq.~\eqref{eq:bound3pure}.

\subsubsection{Proof of Theorem~\ref{thm:bound4}}\label{methods:proof_bound4}

Splitting each term of $w_2$ as $W^2=|W|^{2/3}\,|W|^{4/3}$ and applying H\"older's inequality with conjugate exponents $3/2$ and $3$, we obtain
\begin{align}
w_2&=\sum_{\mathbf u}|W|^{2/3}\,|W|^{4/3}\nonumber\\
&\le \Big(\sum_{\mathbf u}|W|\Big)^{2/3}\Big(\sum_{\mathbf u}W^4\Big)^{1/3}
=\|W\|_1^{2/3}\,w_4^{1/3}. \label{eq:holder4}
\end{align}
Raising both sides to the third power gives $w_2^3\le\|W\|_1^2\,w_4$, which is Eq.~\eqref{eq:bound4}. H\"older's inequality is an equality exactly when the two raised factors are proportional, that is, when $|W|$ is proportional to $W^4$ on the support, which means $|W|$ is constant there.

For product states, the Wigner moments factorize as $w_k(\rho\otimes\sigma)=w_k(\rho)\,w_k(\sigma)$, establishing the additivity of $\mathscr M_4$. Its invariance under Clifford unitaries follows from the permutation covariance of the DWF~\cite{gross_2006}. 

For a pure state $w_2=1/D$, and Lemma~\ref{lem:basic} gives $w_4\le 1/D^3$. Hence $w_2^3/w_4\ge1$ and $\mathscr M_4\ge0$. Equality requires $|W|=1/D$ on the support. The support then has $D$ points, because $w_2=1/D$, and the condition $\sum W=1$ forces all of these values to be positive. The state is then a stabilizer state by the discrete Hudson theorem~\cite{gross_2006}.

\subsection{Multicopy measurement methods}\label{methods:multicopy}
\subsubsection{Proof of Proposition~\ref{prop:Rk}}\label{methods:proof_multicopy}

\begingroup
\renewcommand{\proof}{\par\noindent\textit{Proof.}\enspace}

From the definitions of $T_{\mathbf u}$, $A_{\mathbf 0}$ and Eq.~\eqref{eq:A_u}, the phase-space point operator at $\mathbf u=(\mathbf p,\mathbf q)$ acts as
\begin{align}
A_{(\mathbf p,\mathbf q)}\ket{\mathbf x}=\omega^{2\mathbf p\cdot(\mathbf q-\mathbf x)}\ket{2\mathbf q-\mathbf x}. \label{eq:Aaction}
\end{align}
Applying this to each of the $k$ copies and summing over the phase space,
\begin{align}
&\sum_{\mathbf p,\mathbf q}A_{(\mathbf p,\mathbf q)}^{\otimes k}\ket{\mathbf x_1,\dots,\mathbf x_k}\nonumber\\
&\quad=\sum_{\mathbf q}\Big[\sum_{\mathbf p}\omega^{2\mathbf p\cdot(k\mathbf q-\mathbf s)}\Big]\ket{2\mathbf q-\mathbf x_1,\dots,2\mathbf q-\mathbf x_k}\nonumber\\
&\quad=D\sum_{\mathbf q}\delta_{k\mathbf q,\mathbf s}\ket{2\mathbf q-\mathbf x_1,\dots,2\mathbf q-\mathbf x_k}. \label{eq:sumA}
\end{align}
When $k$ is invertible modulo $d$ the only solution is $\mathbf q=k^{-1}\mathbf s$, so that
\begin{align}
\sum_{\mathbf u}A_{\mathbf u}^{\otimes k}=D\,R_k. \label{eq:sumARk}
\end{align}
Then $w_k=D^{-k}\sum_{\mathbf u}\Tr[A_{\mathbf u}\rho]^k=D^{-k}\Tr[\rho^{\otimes k}\sum_{\mathbf u}A_{\mathbf u}^{\otimes k}]$ gives Eq.~\eqref{eq:wkRk}. The map in Eq.~\eqref{eq:Rk} preserves $\mathbf s$, since the new sum is $2\mathbf s-\mathbf s$. Applying it twice therefore returns $\mathbf x_j$, so $R_k^2=\mathbb I$, and a real permutation matrix that squares to the identity is symmetric.
\endgroup

\subsubsection{Measurement circuit}\label{subsec:circuit}

The operator $R_k$ is a linear map on the labels, so it is a Clifford operation and can be brought to a simple form. Let $V_k$ be the Clifford unitary
\begin{align}
V_k\ket{\mathbf x_1,\mathbf x_2,\dots,\mathbf x_k}=\ket{\mathbf s,\mathbf x_2-\mathbf x_1,\dots,\mathbf x_k-\mathbf x_1}, \label{eq:Vk}
\end{align}
which is invertible because $k$ is invertible modulo $d$, and which is built from two-qudit SUM gates and single-qudit multiplication gates acting on corresponding qudits of different copies. Equation~\eqref{eq:Rk} leaves $\mathbf s$ unchanged and flips the sign of every difference $\mathbf x_j-\mathbf x_1$, so that
\begin{align}
V_kR_kV_k^\dagger=\mathbb I\otimes\Pi^{\otimes (k-1)}, \label{eq:VRV}
\end{align}
where $\Pi=A_{\mathbf 0}$ is the parity operator of one $n$-qudit register. The protocol for one round is therefore as follows. Prepare $k$ copies of $\rho$, apply $V_k$, measure the single-qudit parity on each qudit of the last $k-1$ registers, and multiply the outcomes. The product is a random variable $r=\pm1$ with mean
\begin{align}
\mathbb E[r]=D^{\,k-1}w_k. \label{eq:meanr}
\end{align}
The circuit has depth independent of $n$ and uses only transversal gates between copies. Alternatively, since $R_k$ is both unitary and Hermitian, its expectation value can be obtained from a Hadamard test with one control qubit.

\subsection{Sample complexity of the moment-based criterion}
\label{app:sample_complexity}

Here, we state the concentration inequality used in Sec.~\ref{subsec:samples} and provide the detailed sample-complexity analysis for the moment-based detection and quantification criteria introduced in Secs.~\ref{s3} and \ref{s5}. We use the standard Hoeffding's inequality~\cite{hoeffding_1963} in the form required for our analysis.

\begin{lemma}[Hoeffding]\label{lem:hoeffding}
Let $r_1,\dots,r_M$ be independent random variables with values in $[-1,1]$ and
common mean $\mu$, and let $\bar r=\frac1M\sum_i r_i$. Then, for every $\epsilon>0$,
\begin{align}
\Pr\big[\,|\bar r-\mu|\ge\epsilon\,\big]
\le 2\,e^{-M\epsilon^2/2}.
\label{eq:hoeffdinglemma}
\end{align}
\end{lemma}

Now we prove each part of Corollary~\ref{cor:samples}, one by one.

\subsubsection{Proof of Corollary~\ref{cor:samples}}\label{methods:proof_sampling}

\emph{Part (i).} For the multicopy measurements defined in Sec.~\ref{subsec:circuit}, each round acts on fresh copies of $\rho$, so their outcomes are independent. Each outcome is $\pm1$, and by Eq.~\eqref{eq:meanr} each has mean $\mu=D^{k-1}w_k$. Applying Lemma~\ref{lem:hoeffding}, the probability that the estimate differs from $D^{k-1}w_k$ by $\eta$ or more is bounded by
\begin{align}
\Pr\left[
\left|\frac{1}{M}\sum_{i=1}^{M}r_i
-D^{k-1}w_k\right|\geq\eta
\right]
\leq2e^{-M\eta^2/2}.
\end{align}
This probability is at most $\delta$ when
\begin{align}
e^{-M\eta^2/2}&\leq\frac{\delta}{2},~~
\text{that is,}\quad
\frac{M\eta^2}{2}\geq\ln\frac{2}{\delta}.
\label{eq:solveM}
\end{align}
This establishes Eq.~\eqref{eq:hoeffding}. Since each round consumes $k$ copies of $\rho$, the total copy complexity is $kM$. The bound is independent of $n$ and $d$, as these parameters affect only the expectation value while the measurement outcomes remain bounded.

\emph{Part (ii).} We now derive the copy complexity required to estimate the moment-based detection
criterion
\begin{equation}
    G\equiv P^2-b>0,
    \label{eq:app_G}
\end{equation}
which is equivalent to $w_2^2-w_3>0$, since $G=D^2(w_2^2-w_3)$. Here, $P=Dw_2=\Tr(\rho^2)$ and $b=D^2w_3$ are the rescaled moments defined in Eq.~\eqref{eq:rescaled}. Our objective is to estimate $G$ to additive accuracy $\epsilon$ with confidence at least $1-\delta$.

We consider the multicopy measurement protocol developed in Sec.~\ref{subsec:Rk} for odd prime local dimensions $d\geq5$, where both the second- and third-moment estimators have outcomes in $\{-1,+1\}$.

Let $M_P$ and $M_b$ denote the numbers of independent measurement rounds used to estimate $P$ and $b$, respectively. Each round requires two copies for $P$ and three copies for $b$. Thus, the total number of copies is
\begin{equation}
    N=2M_P+3M_b.
    \label{eq:app_total_copies}
\end{equation}

Let $X_i$ and $Y_j$ denote the measurement outcomes corresponding to the second and third moments, with
\begin{equation}
    \mathbb{E}[X_i]=P,\qquad
    \mathbb{E}[Y_j]=b,
    \qquad X_i,Y_j\in\{-1,+1\}.
\end{equation}
The corresponding sample-mean estimators are
\begin{equation}
    \widetilde P=
    \frac{1}{M_P}\sum_{i=1}^{M_P}X_i,
    \qquad
    \hat b=
    \frac{1}{M_b}\sum_{j=1}^{M_b}Y_j.
    \label{eq:app_estimators}
\end{equation}

Applying Hoeffding's inequality (Lemma~\ref{lem:hoeffding}), we obtain \begin{align}
    \Pr\left[
    |\widetilde P-P|\geq\epsilon_P
    \right]
    &\leq
    2e^{-M_P\epsilon_P^2/2},
    \label{eq:app_hoeffding_P}\\
    \Pr\left[
    |\hat b-b|\geq\epsilon_b
    \right]
    &\leq
    2e^{-M_b\epsilon_b^2/2}.
    \label{eq:app_hoeffding_b}
\end{align}

To ensure that these failure probabilities are at most $\delta_P$ and $\delta_b$, respectively, it is sufficient
to require
\begin{align}
    M_P&\geq
    \frac{2}{\epsilon_P^2}
    \ln\left(\frac{2}{\delta_P}\right),
    \label{eq:app_MP}\\
    M_b&\geq
    \frac{2}{\epsilon_b^2}
    \ln\left(\frac{2}{\delta_b}\right).
    \label{eq:app_Mb}
\end{align}

We next determine how the individual estimation errors propagate to the witness $G$. Since the purity satisfies $0\leq P\leq1$, we define the clipped estimator
\begin{equation}
    \hat P=
    \min\{1,\max\{0,\widetilde P\}\}.
\end{equation}
Clipping cannot increase the estimation error, and hence
\begin{equation}
    |\hat P-P|
    \leq|\widetilde P-P|.
\end{equation}

Defining the estimated witness as
\begin{equation}
    \hat G=\hat P^2-\hat b,
\end{equation}
we obtain
\begin{align}
    |\hat G-G|
    &=\left|
    (\hat P^2-P^2)-(\hat b-b)
    \right|\nonumber\\
    &\leq
    |\hat P-P|\,|\hat P+P|
    +|\hat b-b|\nonumber\\
    &\leq
    2|\hat P-P|+|\hat b-b|,
    \label{eq:app_error_propagation}
\end{align}
where the last inequality follows from $0\leq P,\hat P\leq1$.

Consequently, whenever $|\hat P-P|\leq\epsilon_P$ and $|\hat b-b|\leq\epsilon_b$,
we have
\begin{equation}
    |\hat G-G|
    \leq2\epsilon_P+\epsilon_b.
    \label{eq:app_error_G}
\end{equation}

Using the union bound, the probability that the witness error exceeds this
threshold satisfies
\begin{align}
    &\Pr\left[
    |\hat G-G|>
    2\epsilon_P+\epsilon_b
    \right]\nonumber\\
    &\quad\leq
    \Pr\left[
    |\hat P-P|>\epsilon_P
    \right]
    +
    \Pr\left[
    |\hat b-b|>\epsilon_b
    \right]\nonumber\\
    &\quad\leq\delta_P+\delta_b.
    \label{eq:app_union_bound}
\end{align}

Therefore, an overall additive accuracy $\epsilon$ with failure  probability at most $\delta$ is guaranteed provided that
\begin{equation}
    2\epsilon_P+\epsilon_b\leq\epsilon,
    \qquad
    \delta_P+\delta_b\leq\delta.
    \label{eq:app_constraints}
\end{equation}

Combining Eqs.~\eqref{eq:app_total_copies}, \eqref{eq:app_MP}, and \eqref{eq:app_Mb}, a sufficient total copy budget is obtained by choosing
\begin{equation}
\begin{split}
N={}&
2\left\lceil
\frac{2}{\epsilon_P^2}
\ln\left(\frac{2}{\delta_P}\right)
\right\rceil +
3\left\lceil
\frac{2}{\epsilon_b^2}
\ln\left(\frac{2}{\delta_b}\right)
\right\rceil.
\end{split}
\label{eq:app_general_bound}
\end{equation}

For simplicity, choosing
\begin{equation}
    \epsilon_P=\epsilon_b=
    \frac{\epsilon}{3},
    \qquad
    \delta_P=\delta_b=
    \frac{\delta}{2}
\end{equation}
satisfies the constraints in Eq.~\eqref{eq:app_constraints}. The corresponding sufficient copy budget is
\begin{equation}
    N=
    5\left\lceil
    \frac{18}{\epsilon^2}
    \ln\left(\frac{4}{\delta}\right)
    \right\rceil,
    \label{eq:app_final_bound}
\end{equation}
which guarantees
\begin{equation}
    \Pr\left[
    |\hat G-G|>\epsilon
    \right]\leq\delta.
    \label{eq:app_confidence}
\end{equation}
Equations~\eqref{eq:app_final_bound} and \eqref{eq:app_confidence} are Eqs.~\eqref{eq:Mtot} and \eqref{eq:witness_confidence_main} of part (ii) of Corollary~\ref{cor:samples}.

Thus, the total copy complexity scales as
\begin{equation}
    N=
    O\left(
    \frac{1}{\epsilon^2}
    \log\frac{1}{\delta}
    \right),
    \label{eq:app_scaling}
\end{equation}
independently of the number of qudits for the bounded-outcome implementation.

Finally, the confidence interval $G\geq\hat G-\epsilon$ provides an operational certification condition. In particular, whenever $\hat G>\epsilon$, Wigner negativity is certified with confidence at least $1-\delta$. For a state with a true witness gap $G>2\epsilon$, the protocol detects the violation with probability at least $1-\delta$. Both statements follow from Eq.~\eqref{eq:app_confidence}. If the state is not magic, its DWF is nonnegative, and Theorem~\ref{thm:main} gives $w_2^2\le w_3$, that is, $G\le0$. Declaring it magic requires $\hat G>\epsilon$, and hence
\begin{equation}
\hat G-G>\epsilon-G\ge\epsilon,
\end{equation}
which happens with probability at most $\delta$. Conversely, any estimate below or at the certification threshold, $\hat G\leq\epsilon$, satisfies
\begin{equation}
G-\hat G\geq G-\epsilon>\epsilon.
\end{equation}
Thus, the protocol certifies magic with probability at least $1-\delta$ whenever $G>2\epsilon$. This completes the proof of part (ii) of Corollary~\ref{cor:samples}. 

\emph{Part (iii) and (iv).} To prove part (iii), we apply part (i) to $P$ and to $b$ separately, allowing each a failure probability $\delta/2$. Each estimate then requires
\begin{equation}
M\ge\frac{2}{\eta^2}\ln\frac{2}{\delta/2}=\frac{2}{\eta^2}\ln\frac{4}{\delta}
\end{equation}
rounds of each kind, which uses $2M+3M=5M$ copies. By the union bound, $|\hat P-P|\le\eta$ and $|\hat b-b|\le\eta$ hold together with probability at least $1-\delta$. In part (iv) the same holds for $P$ and $c$, with $2M+4M=6M$ copies. We write $P_-=\hat P-\eta$, $b_+=\hat b+\eta$ and $c_+=\hat c+\eta$, so that on this event $P\ge P_-$, $b\le b_+$ and $c\le c_+$. The argument below uses only these three inequalities, and therefore applies to any confidence intervals of this form. Since $P\ge0$ and $P\ge-b\ge-b_+$ by Eq.~\eqref{eq:rescaledbounds}, the true purity also satisfies
\begin{equation}
P\ge \underline{P}=\max\{0,\,P_-,\,-b_+\}.
\end{equation}
Now the bound of Theorem~\ref{thm:bound3} is the function
\begin{equation}
f(P,b)=\frac{P^2-b}{1+2P+b}=\frac{(1+P)^2}{1+2P+b}-1,
\end{equation}
where the second form follows by adding and subtracting $1+2P+b$ in the numerator. We consider it in the region where $P\ge0$ and $P+b\ge0$. There the denominator is at least $1+P>0$, and the two partial derivatives are
\begin{align}
\frac{\partial f}{\partial b}&=-\frac{(1+P)^2}{(1+2P+b)^2}<0,\\
\frac{\partial f}{\partial P}&=\frac{2(1+P)(P+b)}{(1+2P+b)^2}\ge0,
\end{align}
so $f$ decreases when $b$ grows and does not increase when $P$ decreases. The true point $(P,b)$ lies in this region by Eq.~\eqref{eq:rescaledbounds}. We first raise $b$ to $b_+$ at fixed $P$. Along this step $P+b'\ge P+b\ge0$, so we stay in the region and $f(P,b_+)\le f(P,b)$. We then lower $P$ to $\underline{P}$ at fixed $b_+$. Along this step $P'+b_+\ge \underline{P}+b_+\ge0$, so we again stay in the region and $f(\underline{P},b_+)\le f(P,b_+)$. Together with Theorem~\ref{thm:bound3},
\begin{equation}
f(\underline{P},b_+)\le f(P,b)\le\mathcal N(\rho),
\end{equation}
which is Eq.~\eqref{eq:cert_bound3}. The entry $-b_+$ in $\underline{P}$ is what keeps the second step inside the region. Without it the lower endpoint for $P$ could lie below the line $P+b_+=0$, where $f$ increases as $P$ decreases, and the inserted value could exceed the true bound.

The bound of Theorem~\ref{thm:bound4}, $\sqrt{P^3/c}$, is nondecreasing in $P\ge0$ and decreasing in $c>0$. The true $c=D^3\sum_{\mathbf u}W_\rho(\mathbf u)^4$ is strictly positive, since the Wigner function sums to one. With $P\ge\max\{0,P_-\}\ge0$ and $0<c\le c_+$,
\begin{equation}
\sqrt{\frac{\max\{0,P_-\}^3}{c_+}}\le\sqrt{\frac{P^3}{c}}\le\|W_\rho\|_1
\end{equation}
by Theorem~\ref{thm:bound4}. Using $\|W_\rho\|_1=1+2\mathcal N(\rho)$ from Eq.~\eqref{eq:l1neg}, this is Eq.~\eqref{eq:cert_bound4}. This bound uses only $P$ and $c$, so no information on $b$ is needed. This completes the proof of Corollary~\ref{cor:samples}.

\subsection{Classical shadows of the Wigner function}\label{subsec:shadows}

In each round of the protocol a random unitary $U$ is applied to a fresh copy of $\rho$ and all qudits are measured in the computational basis. If the outcome is $\mathbf k$, the vector $\ket{\phi}=U^\dagger\ket{\mathbf k}$ is stored. When $U$ is drawn from a unitary 2-design, such as the Haar measure or the Clifford group, the operator
\begin{align}
\hat\rho=(D+1)\ket{\phi}\!\bra{\phi}-\mathbb I \label{eq:snapshot}
\end{align}
is an unbiased estimator of the state with $\mathbb E[\hat\rho]=\rho$~\cite{huang_2020}. We call it a snapshot. Its DWF follows from Eq.~\eqref{eq:DWF} and from $\Tr[A_{\mathbf u}]=1$,
\begin{align}
\hat W(\mathbf u)=\frac{(D+1)\bra{\phi}A_{\mathbf u}\ket{\phi}-1}{D}, \label{eq:snapW}
\end{align}
which is again unbiased, i.e., $\mathbb E[\hat W(\mathbf u)]=W_\rho(\mathbf u)$.

A single snapshot does not provide an unbiased estimate of $W_\rho(\mathbf u)^2$, since $\mathbb E[\hat W(\mathbf u)^2]$ also contains the variance of $\hat W(\mathbf u)$. This limitation can be overcome by combining independent snapshots. For $N$ snapshots $\hat W_1,\dots,\hat W_N$ and distinct indices, $\mathbb E[\hat W_i(\mathbf u)\hat W_j(\mathbf u)]=W_\rho(\mathbf u)^2$, and the same holds for products of three and four distinct snapshots. Averaging over all ordered tuples of distinct indices therefore gives unbiased estimators, the so-called U-statistics. They need only the power sums
\begin{align}
S_m(\mathbf u)=\sum_{i=1}^{N}\hat W_i(\mathbf u)^m, \qquad m=1,\dots,4, \label{eq:powersums}
\end{align}
because sums over distinct indices can be written through power sums. With $(N)_k=N(N-1)\cdots(N-k+1)$ the estimators read
\begin{align}
\hat w_2&=\frac{1}{(N)_2}\sum_{\mathbf u}\big[S_1^2-S_2\big], \label{eq:hatw2}\\
\hat w_3&=\frac{1}{(N)_3}\sum_{\mathbf u}\big[S_1^3-3S_1S_2+2S_3\big], \label{eq:hatw3}\\
\hat w_4&=\frac{1}{(N)_4}\sum_{\mathbf u}\big[S_1^4-6S_1^2S_2+8S_1S_3+3S_2^2-6S_4\big]. \label{eq:hatw4}
\end{align}
Each bracket is the sum over distinct tuples, $\sum_{i\neq j}\hat W_i\hat W_j$ for $k=2$ and so on. The violation is estimated as $\hat w_2^2-\hat w_3$. Its bias equals the variance of $\hat w_2$ and becomes negligible once $N$ is large enough for a useful estimate. The certified bounds of Theorems~\ref{thm:bound3} and \ref{thm:bound4} are evaluated with the same estimates.

We would now like to discuss two experimentally relevant implementations of classical shadows. In the global-shadow protocol, the random unitary $U$ acts jointly on all qudits, whereas in the local-shadow protocol, independent random unitaries are applied to individual qudits, offering a simpler experimental implementation. Each local snapshot is a tensor product of single-qudit snapshots, and its DWF factorizes into the corresponding single-qudit Wigner functions. In our numerical simulations, the random unitaries are sampled from the Haar measure, and the states $\ket{\phi}$ are generated according to the probability density $D\bra{\phi}\rho\ket{\phi}$ with respect to this measure. The expectation values $\bra{\phi}A_{\mathbf u}\ket{\phi}$ at all $D^2$ phase-space points are then evaluated using a single fast Fourier transform per snapshot, exploiting the action of the phase-point operators given in Eq.~\eqref{eq:Aaction}.

\subsection{Quasiprobability estimator and its sampling prescription}\label{methods:quasi-simulation}

Consider a system of $n$ qudits initially prepared in a product state $\rho=\rho_1\otimes\dots\otimes\rho_n$, acted on by a Clifford unitary $U$, and measured in the computational basis. We want the probability of an outcome $\mathbf o$,
\begin{align}
p(\mathbf o)=\Tr\big[\,\ket{\mathbf o}\!\bra{\mathbf o}\,U\rho\,U^\dagger\big]. \label{eq:born}
\end{align}
Inserting the expansion of Eq.~\eqref{eq:expansion} for $\rho$ turns this into a sum over phase space,
\begin{align}
p(\mathbf o)=\sum_{\mathbf u}W_\rho(\mathbf u)\,\xi_{\mathbf o}(\mathbf u), \qquad
\xi_{\mathbf o}(\mathbf u)=\bra{\mathbf o}\,U A_{\mathbf u}U^\dagger\ket{\mathbf o}. \label{eq:pquasi}
\end{align}
A Clifford unitary maps each phase-space point operator to another one, $UA_{\mathbf u}U^\dagger=A_{\Gamma(\mathbf u)}$, where $\Gamma$ is an affine map of phase space that is easy to compute~\cite{gross_2006}. A point operator sends $\ket{\mathbf o}$ to $\ket{2\mathbf q-\mathbf o}$ up to a phase, so its diagonal element is one if the position label of the point is $\mathbf o$ and zero otherwise. The function $\xi_{\mathbf o}$ is therefore an indicator,
\begin{align}
\xi_{\mathbf o}(\mathbf u)=\begin{cases}1, & \Gamma(\mathbf u)\ \text{has position}\ \mathbf o,\\ 0, & \text{otherwise}.\end{cases} \label{eq:indicator}
\end{align}
If $W_\rho$ were a probability distribution, Eq.~\eqref{eq:pquasi} would describe a classical stochastic process. One would draw a phase-space point, move it with $\Gamma$, and read off its position. When $W_\rho$ has negative values, one samples from $|W_\rho|$ instead and corrects with the sign. Draw $\mathbf u$ with probability $|W_\rho(\mathbf u)|/\|W_\rho\|_1$ and record
\begin{align}
X=\|W_\rho\|_1\,\mathrm{sign}\big[W_\rho(\mathbf u)\big]\,\xi_{\mathbf o}(\mathbf u). \label{eq:estimator}
\end{align}
By construction, the average of $X$ is exactly $p(\mathbf o)$, and every value of $X$ lies between $-\|W_\rho\|_1$ and $+\|W_\rho\|_1$. For a product input, $|W_\rho|/\|W_\rho\|_1$ is a product distribution, so each sample is drawn qudit by qudit in polynomial time. The sampling budget grows with the Wigner $\ell_1$ norm; implementation and preprocessing costs must also be considered. Averaging $s$ independent values of $X$, Hoeffding's inequality (Lemma~\ref{lem:hoeffding} of Sec.~\ref{subsec:samples}, applied to $X/\|W_\rho\|_1$) gives the estimate within error $\epsilon$ with probability at least $1-\delta$ under the sufficient Hoeffding prescription
\begin{align}
s_{\rm H}(\epsilon,\delta)=\frac{2}{\epsilon^2}\,\|W_\rho\|_1^{2}\,\ln\frac2\delta. \label{eq:pwb}
\end{align}
The Wigner $\ell_1$ norm is multiplicative under tensor products. Thus, when $m$ input qudits are prepared in identical magic states $\rho_{\rm m}$ and the remaining inputs are stabilizer states,
\begin{align}
s_{\rm H}(\epsilon,\delta)=\frac{2}{\epsilon^2}\,\|W_{\rho_{\rm m}}\|_1^{2m}\,\ln\frac2\delta
=\frac{2}{\epsilon^2}\,e^{2m\,\mathscr M(\rho_{\rm m})}\,\ln\frac2\delta. \label{eq:pwbm}
\end{align}
The sufficient sample-count prescription grows exponentially in the number of injected magic states, at a rate set by their mana. This is the estimator of Ref.~\cite{pashayan_2015}, written for Clifford circuits and computational-basis measurements, where all gates and measurements are nonnegative and the input states are the only source of Wigner negativity.
\section{Acknowledgements} 
A.G.M. acknowledges the financial support provided by the IIT Goa Startup Grant (No. 2025/SG/AGM/058).

\bibliography{main}

\section{Supplementary Information}
\noindent This Supplementary Information collects the essential notations, preliminary information and illustrative examples, the numerical data, and the comparison with direct Wigner tomography that support the main text. The proofs of these results are given in the main text and are not repeated here. 

\subsection{Stabilizer Formalism and Magic States} \label{Supp:preliminaries_end}
The stabilizer formalism provides an efficient description of quantum states through commuting subgroups of generalized Pauli operators~\cite{gottesman_1997}. A pure stabilizer state $|\psi\rangle$ is the unique common $+1$ eigenstate of a maximal abelian subgroup of the Pauli group. To formulate this framework for qudit systems, we briefly introduce the generalized Pauli operators, the Heisenberg-Weyl group, and the Clifford group~\cite{veitch_2012}.

For a $d$-dimensional quantum system with computational basis $\{|j\rangle\}_{j=0}^{d-1}$, the generalized Pauli operators are defined as
\begin{equation} 
X|j\rangle=|j+1\;(\mathrm{mod}\;d)\rangle, ~~ Z|j\rangle=\omega^j|j\rangle, 
\end{equation}
where $\omega=e^{2\pi i/d}$ and $ZX=\omega XZ$. The operators $X$ and $Z$, often referred to as the shift and phase operators, respectively, generate the Heisenberg-Weyl group $\mathcal{H}_d$, which describes translations in discrete phase space. For odd $d$, the displacement operator associated with the phase-space point $\mathbf{u}=(p,q)\in\mathbb{Z}_d\times\mathbb{Z}_d$ is
\begin{equation} 
T_{\mathbf{u}} \equiv T_{(p,q)} = \omega^{-(2^{-1}pq)}Z^pX^q, ~~ (p,q)\in\mathbb{Z}_d\times\mathbb{Z}_d, 
\end{equation}
where $2^{-1}=\frac{d+1}{2}$ is the multiplicative inverse of $2$ in $\mathbb{Z}_d$. For $n$ qudits, the Heisenberg-Weyl operators are tensor products of the local ones,
\begin{equation}
T_{\mathbf{u}} = T_{(p_1,q_1)}\otimes T_{(p_2,q_2)}\otimes\cdots\otimes T_{(p_n,q_n)},
\end{equation}
where $\mathbf{u}=(p_1,q_1,\ldots,p_n,q_n)\in\mathbb Z_d^{2n}$ labels the full phase-space point.

The Clifford group $\mathcal{C}_d$ is the normalizer of $\mathcal{H}_d$ in the unitary group, i.e., the set of unitaries that map Heisenberg-Weyl operators to Heisenberg-Weyl operators under conjugation,
\begin{equation}
\mathcal{C}_d=\{U\in\mathrm{U}(d^n)\;|\;UT_{\mathbf{u}}U^{\dagger}\in\mathcal{H}_d\;\;\forall\,T_{\mathbf{u}}\in\mathcal{H}_d\}.
\end{equation}

Clifford operations therefore map stabilizer states to stabilizer states. If $S\in\mathcal{H}_d$ stabilizes $|\psi\rangle$, then for any $U\in\mathcal{C}_d$ the state $|\psi'\rangle=U|\psi\rangle$ satisfies $(USU^\dagger)|\psi'\rangle=US|\psi\rangle=|\psi'\rangle$, and since $USU^\dagger\in\mathcal{H}_d$, $|\psi'\rangle$ is again a stabilizer state.

This closure property underlies the Gottesman-Knill theorem: circuits built from stabilizer-state preparations, Clifford operations, and Pauli measurements, together with classical processing, can be simulated efficiently on a classical computer~\cite{gottesman_1998}. Stabilizer operations alone are thus insufficient for universal quantum computation~\cite{Bravyi_2005}. Universality can be restored by supplementing them with non-stabilizer resources, commonly called \emph{magic states}, which constitute the fundamental resource for quantum computational advantage beyond the stabilizer framework. Motivated by this central role, here we develop an efficient method for certifying magic states. As discussed below, the discrete Wigner representation is particularly well suited to this task, since Wigner negativity is intimately connected to magic in odd and prime power dimensional systems.

\subsubsection{Discrete Wigner Function and Wigner Negativity}
\label{Supp:dwf}
The Wigner function provides a phase-space representation of a quantum state and plays a role analogous to a joint probability distribution over position and momentum in classical statistical mechanics~\cite{wigner_1932}. Unlike a genuine probability distribution, however, it is only a quasiprobability distribution and can take negative values. Such negativity is a signature of nonclassicality, even though the marginals of the Wigner function still reproduce the probability distributions of the corresponding observables. Its finite-dimensional counterpart, the discrete Wigner function (DWF), extends this description to qudit systems~\cite{wootters_1987}. For odd and prime-powered-dimensional systems, the DWF formalism is particularly relevant to the resource theory of quantum magic, as its negativity is closely connected to nonstabilizerness and the limitations of efficient classical simulation~\cite{gross_2006,veitch_2012}.

For a qudit of odd prime dimension $d$, the discrete phase space consists of $d^2$ points $\mathbf{u}=(p,q)\in\mathbb{Z}_d\times\mathbb{Z}_d$. The associated phase-space point operators ${A_{\mathbf{u}}}$ are constructed from the Heisenberg-Weyl displacement operators. Specifically,
\begin{equation}
A_{\mathbf{0}}=\frac{1}{d}\sum_{\mathbf{v}}T_{\mathbf{v}},
\qquad
A_{\mathbf{u}}=T_{\mathbf{u}}A_{\mathbf{0}}T_{\mathbf{u}}^\dagger.
\label{Supp:eq:A_u}
\end{equation}

These operators are Hermitian and form an orthogonal operator basis satisfying
\begin{equation}
\Tr(A_{\mathbf{u}})=1,
\qquad
\Tr(A_{\mathbf{u}}A_{\mathbf{v}})
=d\delta_{\mathbf{u},\mathbf{v}}.
\end{equation}
The DWF of a quantum state $\rho$ is then defined as
\begin{equation}
W_\rho(\mathbf{u})
=\frac{1}{d}\Tr(A_{\mathbf{u}}\rho),
\qquad
\mathbf{u}\in\mathbb{Z}_d\times\mathbb{Z}_d.
\label{Supp:eq:DWF}
\end{equation}
It is real valued and normalized, $\sum_{\mathbf{u}}W_\rho(\mathbf{u})=1$, but need not be nonnegative. Moreover, the state can be reconstructed from its DWF through
\begin{equation}
\rho=\sum_{\mathbf{u}}W_\rho(\mathbf{u})A_{\mathbf{u}}.
\end{equation}
For an $n$-qudit system, these definitions extend naturally to the phase space $\mathbb{Z}_d^{2n}$, with phase-space point operators given by tensor products of the corresponding single-qudit operators.

The connection between Wigner negativity and quantum magic is formalized by the discrete Hudson theorem~\cite{gross_2006}, which establishes that a pure state of odd prime-dimensional qudits has a nonnegative DWF if and only if it is a stabilizer state:
\begin{equation} 
|\psi\rangle\ \text{is a stabilizer state} ~~\Longleftrightarrow ~~ W_\psi(\mathbf{u})\geq0 ~~\forall\,\mathbf{u}. 
\end{equation} 
Consequently, for pure states, $W_\psi(\mathbf{u})<0$ for some $\mathbf{u} ~~\Longleftrightarrow ~~ |\psi\rangle$ is non-stabilizer.

Thus, Wigner negativity provides a necessary and sufficient criterion for nonstabilizerness in pure states. For mixed states, however, the equivalence does not hold: while Wigner negativity certifies magic, a nonnegative DWF does not necessarily imply membership in the stabilizer polytope. These observations motivate the development of experimentally accessible criteria for certifying Wigner negativity without reconstructing the complete quasiprobability distribution. 
\subsection{Illustrative Wigner distributions and moment thresholds}\label{methods:examples}
\subsubsection{Pure-state examples}\label{subsec:examples}
The examples below give the detailed calculations supporting the concise illustrations in Results. The strange-state DWF was defined in Eq.~\eqref{Supp:eq:strange}.
\paragraph{Example: \texorpdfstring{$d=3$}{d=3}} 
(i) We first consider the qutrit strange state~\cite{prakash_2020}
\begin{equation}
|\psi_1\rangle=\frac{\ket{1}-\ket{2}}{\sqrt{2}},
\label{Supp:eq:strange}
\end{equation}
with the DWF 
\begin{equation} 
W_{\psi_1}(p,q)= 
\begin{cases} -\dfrac{1}{3}, & (p,q)=(0,0),\\[1mm] \dfrac{1}{6}, & (p,q)\neq(0,0). \nonumber
\end{cases} \label{Supp:eq:strange_DWF}
\end{equation} 
The corresponding second and third Wigner moments are $w_2=\frac{1}{3}, w_3=0$. Therefore, $w_2^2-w_3 =\frac{1}{9}>0$. Thus, $|\psi_1\rangle$ violates the moment inequality and is therefore certified as a nonstabilizer or magic state.

(ii) As a second example, consider $|\psi_2\rangle =\frac{1}{\sqrt{3}} \left(|0\rangle+|1\rangle+|2\rangle\right)$. Its DWF is given by 

\begin{equation} 
W_{\psi_2}(p,q)= 
\begin{cases} \dfrac{1}{3}, & p=0,\quad q\in\{0,1,2\},\\[1mm] 0, & \text{otherwise}.
\end{cases} 
\end{equation}

Accordingly,  $w_2=\frac{1}{3}, ~~ w_3=\frac{1}{9}$,  which gives $w_2^2-w_3=0$. Hence, $|\psi_2\rangle$ saturates the moment inequality. In fact, $|\psi_2\rangle$ is a stabilizer state, consistent with the prediction of our criterion.

\paragraph{Example: \texorpdfstring{$d=5$}{d=5}}
(i) We next consider representative states in dimension $d=5$. First, let $|\psi_3\rangle =\frac{1}{\sqrt{5}} \left(|0\rangle+|1\rangle+|2\rangle+|3\rangle+|4\rangle\right)$.  The corresponding DWF is
\begin{align}
W_\rho(p,q) = \begin{cases} \frac{1}{5},& p=0,\quad q\in\mathbb Z_5, \\ 0, & \text{otherwise} \end{cases}
\end{align}

The second and third Wigner moments are therefore $w_2=\frac{1}{5}, \qquad w_3=\frac{1}{25}$,  which yields $w_2^2-w_3=0$. Thus, $|\psi_3\rangle$ saturates the moment inequality. Indeed, $|\psi_3\rangle$ is a stabilizer state, consistent with our moment-based criterion.

(ii) As a contrasting example, consider $\ket{\psi_4}$, 
\begin{equation}
\ket{\psi_4}=\frac{\ket0-\ket1+\ket2+\ket3+\ket4}{\sqrt5}.
\label{Supp:eq:psi4}
\end{equation}
Its DWF is 

\begin{align}
W_\rho =
\begin{bmatrix}
\frac{1}{25} & \frac{1}{5} & \frac{1}{25} & \frac{1}{25} & \frac{1}{25} \\[5pt]
\frac{1+\sqrt{5}}{25} & 0 & \frac{1+\sqrt{5}}{25} &
\frac{1-\sqrt{5}}{25} & \frac{1-\sqrt{5}}{25} \\[5pt]
\frac{1-\sqrt{5}}{25} & 0 & \frac{1-\sqrt{5}}{25} &
\frac{1+\sqrt{5}}{25} & \frac{1+\sqrt{5}}{25} \\[5pt]
\frac{1-\sqrt{5}}{25} & 0 & \frac{1-\sqrt{5}}{25} &
\frac{1+\sqrt{5}}{25} & \frac{1+\sqrt{5}}{25} \\[5pt]
\frac{1+\sqrt{5}}{25} & 0 & \frac{1+\sqrt{5}}{25} &
\frac{1-\sqrt{5}}{25} & \frac{1-\sqrt{5}}{25}\\
\end{bmatrix},
\end{align}
where the rows are labelled by $p=0,\dots,4$ and the columns by $q=0,\dots,4$.
In this case, the DWF contains negative phase-space values, indicating its nonclassical character. The corresponding Wigner moments are $w_2=0.2$, $w_3=0.02464$ and hence $w_2^2-w_3=0.01536>0$. The moment inequality is therefore violated, which certifies, by Theorem~1 of the main text, $|\psi_4\rangle$ as a nonstabilizer or magic state.

In all four cases the criterion gives the correct answer, as guaranteed by Theorem~1 of the main text. In terms of the rescaled third moment,
\begin{align}
D^2 w_3=
\begin{cases}
1, & \ket{\psi_2},\ \ket{\psi_3}\ \ (\text{stabilizer}),\\[2pt]
0, & \ket{\psi_1}\ \ (\text{strange state}),\\[2pt]
0.616, & \ket{\psi_4},
\end{cases} \label{eq:examplesb}
\end{align}
so the stabilizer states reach the maximal value $D^2w_3=1$, while both magic states fall strictly below it.
\subsubsection{Mixed-state distributions}
\paragraph{Example: \texorpdfstring{$d=3$}{d=3}}
Here, we first consider the convex mixture of strange state and maximally mixed state of qutrit system, i.e., $\rho_{1}=p|S\rangle\langle S|+\frac{(1-p)}{3}\mathbb{I}_3$ with $p\in[0,1]$. Its DWF is given by
\[W_{\rho_{1}}(\textbf{u}) = \begin{cases} \frac{1}{9}(1-4p),& \;\textbf{u}=(0,0)\; \\ \frac{1}{9}(1+\frac{p}{2}), & \forall\;\textbf{u}\in(\mathbb{Z}_3\times\mathbb{Z}_3)-\{(0,0)\} \end{cases}\]
Accordingly, $w_2=\frac{1}{9}(2p^2+1)$, $w_3=\frac{1}{81}(1-p)(7p^2+p+1)$ and $w_2^2-w_3=\frac{1}{81}p^2(p+2)(4p-1)$. The corresponding witness is plotted in Fig.~1 of the main text.
\paragraph{Example: \texorpdfstring{$d=5$}{d=5}} 

In this particular example, we consider the convex mixture of $|\psi_4\rangle$ (\ref{Supp:eq:psi4}) and maximally mixed state of $d=5$ system, i.e., $\rho_{2}=p|\psi_{4}\rangle\langle \psi_{4}|+\frac{(1-p)}{5}\mathbb{I}_5$ with $p\in[0,1]$. The DWF of this mixed state comes out to be following:
\begin{align}
W_\rho =
\begin{bmatrix}
\frac{1}{25} & \frac{1+4p}{25} & \frac{1}{25} & \frac{1}{25} & \frac{1}{25} \\[5pt]
\frac{1+p\sqrt{5}}{25} & \frac{1-p}{25} & \frac{1+p\sqrt{5}}{25} &
\frac{1-p\sqrt{5}}{25} & \frac{1-p\sqrt{5}}{25} \\[5pt]
\frac{1-p\sqrt{5}}{25} & \frac{1-p}{25} & \frac{1-p\sqrt{5}}{25} &
\frac{1+p\sqrt{5}}{25} & \frac{1+p\sqrt{5}}{25} \\[5pt]
\frac{1-p\sqrt{5}}{25} & \frac{1-p}{25} & \frac{1-p\sqrt{5}}{25} &
\frac{1+p\sqrt{5}}{25} & \frac{1+p\sqrt{5}}{25} \\[5pt]
\frac{1+p\sqrt{5}}{25} & \frac{1-p}{25} & \frac{1+p\sqrt{5}}{25} &
\frac{1-p\sqrt{5}}{25} & \frac{1-p\sqrt{5}}{25}\\
\end{bmatrix}
\end{align}
The second and third order moments of DWF are $w_2=\frac{1}{25}(1+4p^2)$ and $w_3=\frac{1}{3125}(12p^3+60p^2+5)$. Therefore, $w_2^2-w_3=\frac{4p^2}{3125}(20p^2-3p-5)$. 

The criterion certifies the mixed state $\rho_2$ as nonstabilizer for $p\in(\frac{3+\sqrt{409}}{40},1]$. Note that $\frac{3+\sqrt{409}}{40}\approx0.58056$ (see Fig.~1 (b) in the main text).

The noisy strange-state fourth-moment witness is
\begin{equation}
w_2^3-w_4=\frac{p^2(p+2)(4p-1)(4p^2-7p+6)}{1458}.
\label{eq:noisy_strange_fourth}
\end{equation}
Since $4p^2-7p+6>0$, its detection threshold is also $p>1/4$. For the ququint family, the DWF values are $(1+4p)/25$, $1/25$, $(1-p)/25$, $(1+p\sqrt5)/25$, and $(1-p\sqrt5)/25$, with multiplicities $1,4,4,8,8$, respectively. Substituting their power sums into $H_m$ yields Table~1.

\subsection{Comparison with direct Wigner tomography}\label{Supp:tomography}
We now estimate the number of copies required to reconstruct the DWF of an unknown quantum state. Consider a system of $n$ qudits, each of local dimension $d$, with total Hilbert-space dimension $D=d^n$. For odd $d$, the DWF is defined as
\begin{equation}
    W_\rho(\mathbf{u})
    =\frac{1}{D}\Tr[A_{\mathbf{u}}\rho],
    \label{Supp:eq:wigner_tomography_definition}
\end{equation}
where $\mathbf{u}$ labels a point in the discrete phase space containing $D^2$ points. The phase-point operators $A_{\mathbf{u}}$ are Hermitian and satisfy $A_{\mathbf{u}}^2=\mathbb{I}$. Consequently, their measurement outcomes belong to $\{-1,+1\}$, and
\begin{equation}
    -1\leq
    r_{\mathbf{u}}
    \equiv\Tr[A_{\mathbf{u}}\rho]
    \leq1.
\end{equation}

To reconstruct the complete DWF, we consider a direct measurement strategy in which the expectation value of each phase-point operator is estimated independently. Suppose that $m$ copies of $\rho$ are allocated to each phase-space point. The total number of copies required is therefore
\begin{equation}
    N=D^2m.
    \label{Supp:eq:wigner_total_copies}
\end{equation}

Let $\hat r_{\mathbf{u}}$ denote the empirical average of the $m$ measurement outcomes associated with $A_{\mathbf{u}}$. Since the outcomes are independent and bounded in $[-1,1]$, Hoeffding's inequality gives
\begin{equation}
    \Pr\left[
    |\hat r_{\mathbf{u}}-r_{\mathbf{u}}|
    \geq\epsilon
    \right]
    \leq
    2\exp\left(-\frac{m\epsilon^2}{2}\right).
    \label{Supp:eq:wigner_hoeffding}
\end{equation}

The corresponding estimator of the DWF is
\begin{equation}
    \hat W_\rho(\mathbf{u})
    =\frac{\hat r_{\mathbf{u}}}{D}.
\end{equation}
Thus, an additive error $\epsilon$ in the expectation value translates into an error $\epsilon/D$ in the DWF:
\begin{equation}
    |\hat W_\rho(\mathbf{u})
    -W_\rho(\mathbf{u})|
    =\frac{1}{D}
    |\hat r_{\mathbf{u}}-r_{\mathbf{u}}|.
\end{equation}
Accordingly,
\begin{equation}
    \Pr\left[
    |\hat W_\rho(\mathbf{u})
    -W_\rho(\mathbf{u})|
    \geq\frac{\epsilon}{D}
    \right]
    \leq
    2\exp\left(-\frac{m\epsilon^2}{2}\right).
    \label{Supp:eq:wigner_pointwise_error}
\end{equation}

We next require the prescribed accuracy to hold simultaneously at every phase-space point. Define the failure event
\begin{equation}
    B_{\mathbf{u}}
    =
    \left\{
    |\hat W_\rho(\mathbf{u})
    -W_\rho(\mathbf{u})|
    \geq\frac{\epsilon}{D}
    \right\}.
\end{equation}
Applying the union bound over all $D^2$ phase-space points, we obtain
\begin{align}
    \Pr\left[
    \bigcup_{\mathbf{u}}B_{\mathbf{u}}
    \right]
    &\leq
    \sum_{\mathbf{u}}\Pr(B_{\mathbf{u}})
    \nonumber\\
    &\leq
    2D^2\exp\left(-\frac{m\epsilon^2}{2}\right).
    \label{Supp:eq:wigner_union_bound}
\end{align}

To ensure that the probability of at least one estimation error exceeding $\epsilon/D$ is at most $\delta$, it is sufficient to impose
\begin{equation}
    2D^2\exp\left(-\frac{m\epsilon^2}{2}\right)
    \leq\delta.
\end{equation}
Solving for $m$ gives
\begin{equation}
    m\geq
    \frac{2}{\epsilon^2}
    \ln\left(\frac{2D^2}{\delta}\right).
    \label{Supp:eq:wigner_copies_per_point}
\end{equation}

Using Eq.~\eqref{Supp:eq:wigner_total_copies}, a sufficient total copy budget for reconstructing the complete DWF is therefore
\begin{equation}
    N\geq
    \frac{2D^2}{\epsilon^2}
    \ln\left(\frac{2D^2}{\delta}\right).
    \label{Supp:eq:wigner_tomography_bound}
\end{equation}
Here and throughout, integer rounding of the required number of measurement repetitions is understood. This choice guarantees
\begin{equation}
    \Pr\left[
    \max_{\mathbf{u}}
    |\hat W_\rho(\mathbf{u})
    -W_\rho(\mathbf{u})|
    \geq\frac{\epsilon}{D}
    \right]
    \leq\delta.
    \label{Supp:eq:wigner_uniform_confidence}
\end{equation}

For a system of $n$ qudits, substituting $D=d^n$ into Eq.~\eqref{Supp:eq:wigner_tomography_bound} yields
\begin{equation}
    N\geq
    \frac{2d^{2n}}{\epsilon^2}
    \ln\left(\frac{2d^{2n}}{\delta}\right).
    \label{Supp:eq:wigner_nqudit_bound}
\end{equation}
Consequently, the direct pointwise measurement strategy achieves the sample-complexity scaling
\begin{equation}
    N=
    O\left(
    \frac{d^{2n}}{\epsilon^2}
    \ln\frac{d^{2n}}{\delta}
    \right).
    \label{Supp:eq:wigner_tomography_scaling}
\end{equation}

Thus, reconstructing the complete DWF with fixed accuracy in the rescaled phase-space values $D W_\rho(\mathbf{u})$ requires a number of copies that grows exponentially with the number of qudits under this direct measurement strategy. This should be contrasted with the bounded-outcome few-copy moment-estimation protocol, for which the rescaled moment witness can be estimated to fixed additive accuracy using a number of copies independent of $D$, whenever the required bounded-outcome estimators are available.

\end{document}